\documentclass[aps,prb,twocolumn,showpacs,floatfix]{revtex4-1}
\usepackage{graphicx}
\usepackage{dcolumn}
\usepackage{amsmath,mathrsfs,amssymb,amsthm}
\usepackage{hhline}
\usepackage{wasysym,enumerate,color}
\usepackage{epstopdf}
\usepackage{young}
\usepackage{verbatim} 
\usepackage{hyperref}
\hypersetup{colorlinks,linkcolor=blue,urlcolor=blue,citecolor=blue}
\usepackage{color}
\usepackage{ulem} % needed for \scrap-command

\renewcommand{\emph}[1]{\textit{#1}} % needed, if the package  "ulem" is used

\definecolor{darkgreen}{rgb}{0,0.5,0}
\definecolor{purple}{rgb}{0.35,0,0.35}
\definecolor{orange}{rgb}{1,0.5,0}
\definecolor{darkred}{rgb}{.7,0,0}
\definecolor{darkblue}{rgb}{0,0,.3}
\definecolor{grey}{rgb}{.6,.6,.6}
\definecolor{dimgreen}{rgb}{0.2,0.6,0.1}
\hypersetup{colorlinks,linkcolor=blue,urlcolor=blue,citecolor=blue}

\newcommand{\scrap}[1]{{\color{grey}{\sout{#1}}}}
\newcommand{\Sec}[1]{Sec.~\ref{#1}}

\newcommand{\hc}{{\rm h.c.}}
\newcommand{\be}{\begin{equation}}
\newcommand{\ee}{\end{equation}}
\newcommand{\bea}{\begin{eqnarray}}
\newcommand{\eea}{\end{eqnarray}}

\newcommand{\G}{{\Gamma}}
\newcommand{\g}{{\gamma}}
\newcommand{\uG}{\Gamma}
\newcommand{\ug}{\gamma}
\newcommand{\ds}{\displaystyle}

\newcommand{\tr}{{\rm Tr\;}}
\newcommand{\e}{\varepsilon}
\newcommand{\ua}{\alpha}
\newcommand{\ub}{\beta}
\newcommand{\ti}{{\tilde i}}
\newcommand{\tj}{{\tilde j}}
\newcommand{\D}{{\Delta}}

\usepackage{braket}

\usepackage{kbordermatrix}
\renewcommand{\kbldelim}{(}
\renewcommand{\kbrdelim}{)}

\usepackage[caption=false]{subfig}
\usepackage[vcentermath]{youngtab}

\newcommand{\SU}[1]{\ensuremath{\text{SU}(#1)}}

\newcommand{\GTP}{GT-pattern}

\begin{document}
\title{\SU 3 Anderson impurity model: A numerical renormalization 
group approach exploiting non-Abelian symmetries}
\author{C\u at\u alin Pa\c scu Moca}
\affiliation{
BME-MTA Exotic Quantum Phase Group, Institute of Physics, Budapest University of Technology and Economics,
H-1521 Budapest, Hungary\\
Department of Physics, University of Oradea, 410087, Oradea, Romania}
\author{Arne Alex}
\affiliation{Physics Department, Arnold Sommerfeld Center for Theoretical Physics and Center for NanoScience, Ludwig-Maximilians-Universit\"at M\"unchen, D-80333 M\"unchen, Germany}
\author{Jan von Delft}
\affiliation{Physics Department, Arnold Sommerfeld Center for Theoretical Physics and Center for NanoScience, Ludwig-Maximilians-Universit\"at M\"unchen, D-80333 M\"unchen, Germany}
\author{Gergely Zar\'and}
\affiliation{BME-MTA Exotic Quantum Phase Group, Institute of Physics, Budapest University of Technology and Economics,
H-1521 Budapest, Hungary}
\date{\today}
\begin{abstract}
We show how the density-matrix numerical renormalization group (DM-NRG)
method can be used in combination with non-Abelian symmetries such as \SU N, the decomposition of the direct product of two irreducible representations requires the use of a so-called outer multiplicity label. We apply this scheme 
to the \SU 3 symmetrical Anderson model, for which we analyze the finite size spectrum, 
determine local fermionic, spin, superconducting, and trion spectral functions, and also compute the temperature dependence of the conductance. Our calculations reveal a rich Fermi liquid 
structure. 
\end{abstract}
\maketitle
\section{Introduction}

Quantum impurity models, describing a quantum system with a small
number of discrete states, the impurity, coupled to a continuous
bath of fermionic or bosonic excitations, arise in a variety of
contexts. A particularly important example is the Anderson impurity
model,~\cite{Anderson1961} relevant for describing magnetic moments in
metals, transport through quantum dots, and for the treatment of
correlated lattice models using dynamical mean field theory.
While the standard version of this model has \SU 2 spin symmetry in the
absence of a magnetic field,  generalizations to settings with
higher symmetry have also been studied.
 The \SU N generalization of the Anderson model emerged
first in the context of heavy Fermion systems,\cite{Bickers} where
large $N$ expansions proved to be an efficient way to model and
describe magnetic atoms with orbital degeneracy.  Studying these
models in detail is not only useful in the context of heavy fermion
systems, but it also represents the first step to understand the
behavior of correlated cold atomic gases with \SU N symmetrical
interactions. \cite{Honerkamp:2004hq, Rapp:2007gl}

The \SU N Anderson model can also be realized in a controlled way. In
particular, the \SU 4 model has been realized in various mesoscopic
structures including carbon nanotubes,\cite{Kouwenhoven,Finkelstein}
vertical quantum dots,~\cite{Tarucha} and more recently in the
originally proposed double dot
structures.~\cite{Davidunpublished} Similarly, the \SU 3 Anderson
model could also be realized with quantum dot structures, though the
proposed set-up is maybe somewhat more complicated. \cite{Carmi}

The \SU N Anderson model is defined in terms of $N$ local orbitals
embedded in a conduction electron sea. Its Hamiltonian can be written
in terms of the corresponding creation operators,
$d^{\dagger}_{\alpha}$ ($\alpha=1,\dots,N$) and the number operator
$\hat n= \sum_{\alpha} d^{\dagger}_\alpha d_\alpha$ as \bea H &=&
\e_d\;\hat n + \frac U 2 \hat n (\hat n-1) + V \sum_\alpha
(d^{\dagger}_{\alpha} \psi_\alpha(0)
+ \hc) \nonumber \\
&& \quad+\; H_{\rm chan}[\psi,\psi^\dagger]\;.
\label{eq:H_SUN}
\eea Here $\e_d$ and $U$ denote the position of the local orbital and
the strength of interaction on it, respectively, and the level
$d^\dagger$ hybridizes locally with the fermions at its position,
destroyed by $\psi_\alpha(0)$. The last term of the Hamiltonian
describes the kinetic energy of the conduction electrons. It generates
the dynamics of the field $\psi_\alpha(0)$, and amounts in a
broadening of the "atomic" level, $\e_d$.

In the present paper we show on the prototypical example of the
  \SU 3 symmetrical Anderson model how the numerical renormalization
  group\cite{Wilson,Bulla} (NRG) of Ken Wilson, one of the most
  versatile and reliable tools for treating quantum impurity models,
  can be adapted to fully take advantage of non-Abelian symmetries to
  reduce computational costs.  Within Wilson's procedure, one
rewrites \eqref{eq:H_SUN} as Hamiltonian of a semi-infinite chain,
and diagonalizes it iteratively.~\cite{Bulla} Using symmetries in the
course of this diagonalization procedure is crucial: it allows
computer memory to be used efficiently, and enables one to reach
the required numerical accuracy on relatively standard computers with
reasonable runtimes. Eq.~\eqref{eq:H_SUN} obviously possess a
$\SU N \times \text{U}(1)$ symmetry corresponding to rotations in spin space
and overall charge conservation. Here, we shall focus on the $N=3$
case, classify states and observables while exploiting these
symmetries, and determine the spectral functions of several local
observables.

In an earlier work, a general framework has been set up and
implemented to handle an arbitrary number of non-Abelian symmetries
dynamically.\cite{Toth} This formulation allowed us to build an
open-access flexible density-matrix NRG (DM-NRG)
code.\cite{OpenAccess} However, in Ref.~\onlinecite{Toth} we considered only
combinations of certain rather simple symmetries such as charge
and spin \SU 2 symmetries, $\text{Z}_2$ or $\text{U}(1)$ symmetries.  In a group
theoretical sense, these are simpler than \SU{N>2} and some other
discrete or Lie groups. For \SU 2, irreducible representations (irreps)
are labeled by the size of the spin.  
When "adding" two \SU 2 spins, say $S_1$ and $S_2$, each possible 
total spin $S_3$ that satisfies the angular momentum addition rule  
$|S_1 - S_2| \le S_3 \le S_1 + S_2$ can be obtained in precisely one
way. More technically, in the decomposition of the direct product
of two \SU 2 irreps labeled by spins $S_1$ and $S_2$ into a
direct sum of irreps, the number of times $n_{S_1 S_2; S_3}$ that
any irrep labeled by spin $S_3$ occurs in the direct sum, the so-called 
``outer multiplicity'', is either 0 or 1.  This is, however, not
true for \SU{N>2}: the decomposition of the direct product
$\Gamma_1 \otimes \Gamma_2$ of two \SU N irreps into a direct sum
can contain irreps with outer multiplicity $n_{\G_1 \G_2 ;\G_3}$
larger than 1, in other words, there may be $n_{\G_1 \G_2 ; \G_3}$
inequivalent ways to construct the irrep $\Gamma_3$ from $\G_1$ and
$\G_2$.  Correspondingly, the Clebsch-Gordan coefficients have a
more complicated structure in this case, and the Wigner-Eckart
theorem, extensively used in the DM-NRG calculations, becomes also
somewhat more complicated. Here we show how a general framework can be
constructed to deal with this case,\cite{footnote0} and demonstrate it
on the specific example of the \SU 3 Anderson model.

We note that another general approach towards exploiting non-Abelian
symmetries such as \SU N or Sp($N$) within NRG, and more generally for
tensor network methods, has recently been published by
A. Weichselbaum.~\cite{Andreas} It is formulated in the language of
matrix product states, and may be regarded as complementary to our
own, which is phrased within the more traditional formulation of
NRG. We emphasize, though, that both approaches are fully
  equivalent, in that precisely the same NRG assumptions and
  approximations are made in both; they differ only in the data
  structures used for internal book-keeping in the numerical
  codes. Their relation is briefly scetched in App.~\ref{append:MPS}.

Our DM-NRG calculations require explicit knowledge of Clebsch-Gordan
coefficients. Whereas these are known in closed form for \SU 2, this
is not the case for \SU{N>2}. However, an efficient numerical
algorithm for their evaluation has recently been
developed,~\cite{ArneJan} which we use here.

This paper is structured as follows: In \Sec{sec:nonAbelianSymmetries}
we outline our approach for exploiting non-Abelian symmetries in
DM-NRG calculations. In \Sec{sec:SU3-Anderson} we apply it to the \SU 3
Anderson model; in particular, we present results for the
conductance through a quantum dot described by this model, and for
various local spectral functions. Our conclusions in
\Sec{sec:conclusions} are followed by four appendices, that summarize
some basic facts of \SU N representation theory, and some recursion
formulas involving Clebsch-Gordan coefficients, respectively.

\section{DM-NRG with non-Abelian symmetries}
\label{sec:nonAbelianSymmetries}

As stated in the introduction, the formalism presented in
Ref.~\onlinecite{Toth} applies only for a special (though relatively
large) class of symmetries. Therefore, let us review here the most
important formulas and the structure of the NRG calculations for the
more general case, where so-called outer multiplicities are also
considered. \cite{typos}

%We shall, however, not present here all rather lengthy and
%involved formulas, which shall be presented only as accompanying
%supplementary information.\cite{EPAPS}

\subsection{Local symmetries on the Wilson chain}

Let us start by first discussing the general structure of the symmetries of the Wilson 
chain. The first step in Wilson's procedure of solving a quantum impurity problem is to 
perform a Gram-Schmidt orthogonalization  and rewrite  the Hamiltonian in a ``tridiagonal''
form,\cite{Wilson} 
\be
{H} = {\cal H}_{0} +\sum_{n=0}^{\infty}\bigl( \tau_{n, n+1} +{\cal H}_{n+1}\bigr) \;. 
\label{eq:H_Wilson_chain}
\ee
Here ${\cal H}_{0}$ contains the local, interacting part of the Hamiltonian, while the rest of the chain represents the conduction electron (bath) degrees of freedom, coupled to it. 
The on-site terms ${\cal H}_{n+1}$ are many times missing; they typically appear for more sophisticated electronic densities of states and can also account for superconducting correlations.
In our case, one could take, e.g.,  
\bea
{\cal H}_{0}  &=& \e_d\;\hat n + \frac U  2 \hat n (\hat n-1) + \tilde V 
\sum_\alpha (d^{\dagger}_{\alpha} f^{[0]}_\alpha
+ \hc) \;,
\label{Wilson_H0}
\eea 
as ${\cal H}_{0}$, with $f^{[0]}_\alpha\sim\psi_\alpha(0)$
a properly normalized on-site fermion,  and the hopping terms $\tau_{n, n+1}$ would read, 
\be
\tau_{n, n+1}= \sum_{\alpha} h^{[n]} \left(
 f^{[n]\dagger}_{\alpha} f^{[n+1]}_{\alpha} +\hc \right)\;.
\label{Wilson_Tau}
\ee
Here $ f^{[n+1]}_{\alpha}$ annihilates  a fermion of \SU 3 spin $\alpha$ at site $[n]$, and 
 the hopping amplitudes $h^{[n]}$ decay exponentially along chain, 
thereby leading to energy scale separation.  Eq.~\eqref{eq:H_Wilson_chain} is 
then diagonalized iteratively using the recursive relation
\be
H_n= H_{n-1}+\tau_{n-1,n} + {\cal H}_n\;.
\label{eq:H_Wilson_recursive}
\ee

In the following, we shall assume that $H$ (and $H_n$) are invariant
under the direct product of $n_S$ symmetry groups, \be G = {\cal
  G}_{1}\times{\cal G}_{2}\times\dots\times{\cal G}_{n_S}\,.  \ee This
means that for any group element $g_{\lambda}\in {\cal G}_{\lambda}$,
with $\lambda = 1, \dots, n_S$, there exists a unitary operator
${\cal U}_\lambda (g_{\lambda})$ on the Fock space, leaving $H$
invariant.  Here we do not need to make much restriction on the
groups: our considerations hold for any group which acts on the chain
locally at each lattice site $n$ 
\be {\cal U}(g) =
\prod_{\lambda=1}^{n_S} {\cal U}_\lambda (g_{\lambda}) =
\prod_{\lambda=1}^{n_S}\prod_n {\cal U}_{\lambda,n} (g_{\lambda}) \,,
\ee 
and for which the Wigner-Eckart theorem holds.\cite{footnote1} Given
the above group structure, we can group all eigenstates and also
operators into {\it multiplets}, each of which transforms according to a certain
representation of $\cal G$, 
\be \uG \equiv \left\{\G^1,
  \G^2,...,\G^{n_S}\right\}\leftrightarrow \G^1\otimes \G^2\otimes
\dots \otimes\G^{n_S}\;.  \ee 
The "quantum numbers" $\uG^\lambda$, which label the
various irreducible representations (irreps) occurring in $\Gamma$, can
be spin labels, charges, or can label different 
irreps of some point group.  States within a multiplet are
then labeled by a set of internal indices, $\ug =\left\{ \g^1,
  \g^2,...,\g^{n_S} \right\}$, with the internal labels running from
$1\le \g^\lambda\le{\rm dim }(\G^\lambda)$. A given multiplet $i$ 
that transforms according to the representation $\Gamma_i$ thus
consists of \be \dim (i)\equiv\dim
(\uG_i)=\ds{\prod_{\lambda=1}^{n_S}}\dim (\G_i^{\lambda})
\label{eq:dim_i}
\ee
degenerate states.

States belonging to a product of two representations,
$\G_1^\lambda\otimes \G_2^\lambda$, often appear in the
calculations. As outlined in the introduction, similar to spins, these
can be decomposed into irreps. 
However, one irrep
may appear several times in this
decomposition,
\be
\G_s^\lambda\otimes \G_p^\lambda \to ... \oplus 
{\underbrace{\G_q^\lambda \oplus ...\oplus \G_q^\lambda}_{
n_{s,p;q}^{\lambda}
%n_{\G_s^\lambda,\G_p^\lambda}^{\G_q^\lambda} 
\mbox{ times}}}
\oplus ...\phantom{n}.
\label{eq:product_decomposition}
\ee
Accordingly, in the most general case, the resulting states 
$ \G_s^\lambda\otimes \G_p^\lambda \to \Gamma_q^\lambda$ must be labeled by a so-called "outer multiplicity" label, $\alpha_\lambda = 1,.., n_{s,p;\;q}^{\lambda}$. Correspondingly, 
properly transforming multiplets  may be constructed from  
the product of two multiplets as
\begin{equation}
| \uG, \g\rangle_\alpha = 
\sum_{\lambda_1,\lambda_2}\left (   \G_1,\ug_1; \G_2,\ug_2| \G,\ug   \right )_{\ua }\;
|\G_1,\ug_1\rangle\otimes |\G_2,\ug_2 \rangle\;,
\label{eq:decomposition1}
\end{equation}
where $\ua\equiv\{\ua_\lambda\}$ denotes the composite multiplicity label, and the generalized Clebsch-Gordan coefficients are defined as
\be
\left ( \uG_1, \ug_1; \uG_2 ,\ug_2 \mid  \uG,\ug   \right )_{\ua }
\equiv \prod_{\lambda=1}^{n_S}\left ( \uG_1^{\lambda},
\ug_1^{\lambda}; \uG_2^{\lambda} ,\ug_2^{\lambda} \mid  \uG^{\lambda},\ug^{\lambda} \right )_{\alpha_{\lambda} }\;.
\label{eq:general_Clebsch}
\ee

The outer multiplicity also appears in the Wigner-Eckart theorem. The
latter states that the matrix elements of an operator multiplet,
i.e. a set of operators $\{A_{\uG_A,\ug_A}\}$ transforming under
transformations ${\cal U}(g)$ as a multiplet $\uG_A$, are determined
almost entirely by representation theory, and can be expressed in
terms of the Clebsch-Gordan coefficients as
\bea
&&\langle i, \uG_i, \ug_i  |  A_{\uG_A,\ug_A}    | j, \uG_j, 
\ug_j  \rangle = 
\nonumber\\
&&\quad\quad\sum_{\ua}  \left ( \uG_A,
\ug_A; \uG_j ,\ug_j \mid  \uG_i,\ug_i   \right )^{\ast}_{\ua }\langle i\parallel A 
\parallel j\rangle _{\ua}\;.
\label{eq:WE}
\eea
Here multiplets $i$ and $j$ transform according to the
irreps, $\uG_i$ and $\uG_j$.
Thus, according to the Wigner-Eckart theorem, all matrix elements are
determined by only a few reduced matrix elements, $\langle i\parallel
A \parallel j\rangle _{\ua}$, labeled just by the outer multiplicity
labels, $\alpha$ characterizing how many times the representation
$\uG_i$ appears in the product of $\uG_j$ and $\uG_A$. For many
commonly used symmetries as \SU 2, e.g., the outer multiplicity is
just always one and the label $\alpha$ can be dropped. However, it is needed
for, e.g., \SU{N\ge 3} or even for cubic point groups.

\subsection{Wilson's NRG with symmetries}

In course of the NRG procedure, one diagonalizes
Eq.~\eqref{eq:H_Wilson_chain} iteratively. The eigenstates of the
Hamiltonian $H_n$ of a chain of length $n$ can be grouped into
multiplets, with each multiplet $i$ transforming according to a
certain representation $\uG_i=\{\uG_i^1,..,\uG_i^{n_S} \}$.  Having
computed the approximate eigenstates (\emph{block states}) $|i, \G_i,
\g_i, \rangle^{[n-1]}$ of $H_{n-1}$, one proceeds to construct
eigenstates of $H_{n}$.  To do that, one first appends to the chain
the multiplets $\{|\mu, \G_\mu^{\rm loc}, \g_\mu^{\rm loc} \rangle \}
$, spanning \emph{local} Hilbert space at site $n$, and then
constructs properly transforming multiplets $\{| u, \uG_{u},
\ug_{u}\rangle^{[n]}_{\ua_{u}}\}$ by making use of the Clebsch-Gordan
coefficients, Eq.~\eqref{eq:decomposition1}:

 \be | \mu, \uG_\mu^{\rm loc}, \ug_\mu^{\rm loc}\rangle
\otimes | i, \uG_i,\ug_i \rangle^{[n-1]}\to | u, \uG_{u},
\ug_{u}\rangle^{[n]}_{\ua_{u}}\;.  \ee
%\bea
%\left | u, \uG_{u}, \ug_{u}\rangle^{[n]}_{\ua_{u}} & = & \sum_{\ug_i, \ug_\mu^{\rm loc}}\left 
%( \uG_\mu^{\rm loc},\ug_\mu^{\rm loc}; \uG_i ,\ug_i \mid  \uG_{u},\ug_{u}   \right )_{\ua_{u}}\times \nonumber\\ & 
%& \left | \mu, \uG_\mu^{\rm loc}, \ug_\mu^{\rm loc}\rangle 
%\otimes \left | u, \uG_u,\ug_u    \rangle ^{[n-1]}.
%\eea
Notice that a new multiplet  $u$ now also carries an outer multiplicity label, $\alpha_u$: This specifies the representation according to which $\uG_u$ has been produced from $\uG_i$ and $\uG_\mu$. 
The advantage of using these states is that $H_n$ is diagonal both in $\uG_u$ and in the internal 
labels, $\gamma_u$.  Therefore, it is sufficient to compute only the corresponding 
irreducible matrix elements  $\langle u \parallel H_n\parallel v\rangle\big.^{[n]}$ in each 
symmetry sector (block) separately, and diagonalize $H_n$ sectorwise by a unitary transformation
to obtain the  corresponding new eigenstates, 
\be
| u, \uG_{u}, \ug_{u}\rangle ^{[n]}_{\ua_{u}}  \to | \ti, \G_{\ti}, \g_{\ti}  \rangle^{[n]}\;.
\label{eq:transformation}
\ee 
 As explained in App.~\ref{append:MPS}, this iterative procedure leads to a matrix product state (MPS) with a peculiar structure, reflecting the symmetry of the Hamiltonian. 
%\bea
%&&Clebsch\left | \ti, \G_{\ti}, \g_{\ti}  \rangle^{[n]} = \sum_{u,\gamma_u} {\cal O}_{u,\ti}^{[n]} \left | u, 
%\G_u, \g_u \rangle_{\alpha_u }^{[n]}\; \delta_{\G_{\ti}, \G_u}\;\delta_{\g_{\ti},\g_u}, 
%\nonumber
%\label{eq:canonical_transformation}
%\\
%&&
%\bigl< \ti,\;  \uG_{\ti} ,\ug_{\ti}\; 
%|\;H_n\;|\; \tj,\;  \uG_{\tj}, \ug_{\tj}
%\bigr>^{[n]} =E_{\ti}^{[n]}\;
%\delta_{\uG_{\ti},\uG_{\tj}} 
%\delta_{\ug_{\ti},\ug_{\tj}}\;.
%\nonumber
%\eea

The most difficult part in the procedure above is to determine  the matrix elements 
$\langle u \parallel H_n\parallel v\rangle\big.^{[n]}$. These can be constructed by noticing that 
each state $u$ and $v$ has been constructed from the eigenstates of $H_{n-1}$ and ${\cal H}_n$, 
$i,\mu\to u$  and $j,\nu\to v$, and therefore 
 \be
 \langle u \parallel H_n\parallel v\rangle^{[n]} = \delta_{u,v}(E^{n-1}_i + \epsilon^{n}_\mu)
 + \langle u \parallel \tau_{n-1,n}\parallel v\rangle^{[n]} , %\vspace{1mm}
 \nonumber
 \ee 
with $\epsilon^{n}_\mu$ being the  eigenenergy of ${\cal H}_n$. The matrix elements 
of $\tau_{n-1,n}$ can be worked out by assuming  that the hopping part
consists of some fermionic or bosonic \emph{creation} operators $ C_{a,\G_a,\g_a}^{[n]} $, 
transforming again according to some 
irreps $\G_a$, 
\be
\tau_{n-1, n}= \sum_{a}
\left [    
h_{a}^{[n-1]}\sum_{\ug_a}\; C_{a,\uG_a,\ug_a}^{[n-1]}(C_{a,\uG_a, \ug_a}^{[n]})^\dagger +\hc
\right ].
\label{eq:hopping}
\ee Here $a$ labels the different "hopping operators", and
$h_{a}^{[n-1]}$ the corresponding hopping amplitudes. Notice the
somewhat unusual way this hopping term is written:
$C_{a,\uG_a,\ug_a}\leftrightarrow f^\dagger$ is a "creation operator",
which transforms according to the representation $\uG_a$, while
$(C_{a,\uG_a,\ug_a})^\dagger\leftrightarrow f$ is an "annihilation
operator", transforming according to the conjugate representation,
${\uG_a}^*$.  We remark that for
charge \SU 2 symmetry, e.g., the "creation" operator multiplet
$C_{a,\uG_a,\ug_a}$ is a Nambu spinor, and contains both $f$ and
$f^\dagger$ operators.\cite{Bulla1998} The number of hopping
operators may depend on the symmetry used: for a chain of spin 1/2
fermions treated in terms of $\SU 2 \times \text{U}(1)$ symmetry, e.g., one
has a single hopping operator of spin 1/2 and charge 1, while if only
the charge symmetry is used then one has two hopping operators of
charge 1, corresponding to the spin up and spin down directions.  In
our example of the \SU 3 Anderson model we have a single hopping
operator, and $C_{\uG_a,\ug_a}^{[n]} \leftrightarrow \{
f^{[n]\dagger}_{\alpha}\}$. 
 Assuming then that the reduced matrix elements of 
 the creation operators acting on site 
 $n-1$ of the chain, 
$ \langle u \parallel C^{[n-1]}_a \parallel v\rangle_{\ua}^{[n-1]}$,
   and those of the local creation operators at the added site, 
   $\langle\nu\parallel C^{[n]}_a\parallel \mu\rangle_{\ub} $
  are known, one can use the Wigner-Eckart theorem to express  $\langle u \parallel \tau_{n-1,n}\parallel v\rangle^{[n]}$  as
 
 \begin{widetext} 
\begin{multline}
\langle u\parallel\tau_{n-1,n}\parallel v\rangle\big.^{[n]}
=\delta_{\uG_{u},\uG_{v}} \ds {\sum_{a,\ua, \ub}}h_a^{[n-1]}\;
\langle i\parallel C^{[n-1]}_a \parallel j\rangle_{\ua}^{[n-1]}\; 
\langle\nu\parallel C^{[n]}_a\parallel
\mu\rangle_{\ub}^{\ast}\; D\left(a,  \ua, \ub; u, v \right)
  + \hc\;, \phantom{nn}\\
  (u\leftarrow\mu,i; \; v\leftarrow \nu,j)
\;.
\end{multline}
\end{widetext}
 Here the outer multiplicity labels $\alpha$ ($\beta$) label inequivalent ways in which $\G_i$ ($\G_\nu$)
appear in the product  $\G_a\otimes \G_j$ ($\G_a\otimes \G_\mu$). The coefficients
$D\left(a,  \ua, \ub; u, v \right)$ can be expressed in terms of Clebsch-Gordan coefficients, and are 
given by Eq.~\eqref{eq:D} in Appendix~\ref{recursion_formulas}.
Similar expressions  hold for the matrix elements of "block operators" ($A$), acting somewhere  on 
the first $n-1$ sites of the chain, and those of "local operators"   $A^{[n]}$, 
acting only on the last site of the  chain.\cite{footnote2} For a block operator we 
have, e.g., 
%\begin{widetext}
  \begin{equation}
\langle u \parallel A \parallel  v \rangle ^{[n]}_{\ub} =
\sum_{\ua} 
\langle i \parallel A \parallel   j \rangle^{[n-1]}_{\ua}\;
 F\left (\ua, \ub; u, v \right )\;\delta_{\mu, \nu}\; ,
 \label{eq:block_operator}
\end{equation}
while for the local operators the following equation holds, 
\begin{equation}
\langle u \parallel A^{[n]} \parallel  v\rangle^{[n]}
_{\ub} =
\sum_{\ua} 
\langle \mu \parallel A^{[n]} \parallel  \nu \rangle _{\ua}
\;
 K\left (\ua, \ub; u,v \right )\;\delta_{i, j}\;,
\label{eq:local_operator}
\end{equation}
%\end{widetext}
with the coefficients $F\left (\ua, \ub; u, v \right )$ and $K\left (\ua, \ub; u,v \right )$
given in Appendix~\ref{recursion_formulas}.  Here again, the outer multiplicity labels $\beta$  label inequivalent ways in which $\G_u$  appears in $\G_A\otimes \G_v$, while $\alpha$
labels similarly inequivalent  ways how $\G_i $ ($\G_\mu$) can be constructed 
from   $\G_A$ and $\G_j$ ($\G_A$ and $\G_\nu$). 
Similar to $D$,
the coefficients
$F$ and $K$ are again determined only by symmetry, and can be expressed in terms of Clebsch-Gordan coefficients. As a last step of the iteration, the reduced matrix elements 
$\langle u \parallel A^{[n]} \parallel  v\rangle^{[n]}_{\ub}$ and 
$\langle u \parallel A \parallel  v\rangle^{[n]}_{\ub}$ need be transformed to the new basis, 
$|\ti, \G_\ti, \g_\ti \rangle $. This is performed by using precisely the same 
unitary block transformations as the ones used to  diagonalize the Hamiltonian $H_n$, 
\eqref{eq:transformation}, without affecting the outer multiplicity labels $\ub$.

Wilson's diagonalization procedure can be carried out then based upon the 
equations above: In a given iteration, one 
takes the lowest lying states of iteration $[n-1]$ and their matrix elements  $\langle i \parallel C_a^{[n-1]}\parallel j\rangle_\beta^{[n-1]}$, computes from these and from the matrix elements $\langle \mu \parallel C_a^{[n]}\parallel \nu\rangle_\beta$ the Hamiltonian $\langle u \parallel H_n\parallel v\rangle^{[n]}$. Then diagonalizing $\langle u \parallel H_n\parallel v\rangle^{[n]}$,  one obtains low-lying eigenstates of $H_n$ and determines their matrix elements  $\langle \ti \parallel C_a^{[n]}\parallel \tj\rangle_\beta^{[n]}$.
Continuing this procedure for larger and larger values $n$, one obtains better and better 
approximations for the ground state of $H=H_{n\to\infty}$ and the lowest lying eigenstates.

\subsection{FDM-NRG approach}

So far, we discussed essentially Wilson's original NRG approach in
case of general symmetries. In practice, however, one often needs to
go beyond Wilson's RG and use the so called DM-NRG method,\cite{Hofstetter2000}
whereby a forward NRG run is first performed to obtain the density
matrix (DM) of the system, and then a backward NRG run is made to
compute physical observables. Moreover, to satisfy spectral sum rules,
a complete basis set\cite{AndersSchiller2005,Anders}  has to be used, 
as first implemented in the context of DM-NRG
in Refs.~\onlinecite{Peters2006,Weichselbaum2007}. 
In the full density-matrix NRG approach
(FDM-NRG) of Ref.~\onlinecite{Weichselbaum2007}, the
full density matrix of the entire chain is expressed
in the complete basis, which yields an improved treatment of
finite-temperature properties. Let us now
briefly discuss how symmetries can be implemented in the FDM-NRG 
approach. (For a complimentary formulation of the same strategy
using matrix product states, see Ref.~\cite{Andreas}.)

First, to satisfy the necessary completeness relations, we consider a
chain of $N$ sites and introduce "environment" states $e$ for a each
state discarded in iteration $[n]$ ($i\in
D$),\cite{Peters2006,Weichselbaum2007} \be |i, \uG_{i}\; \ug_{i}
\rangle^{[n]}\rightarrow |i, \uG_{i}\; \ug_{i}; e_n \rangle^{[n]}\;.
\label{complete}
\ee
Here the states $e_n$ form an orthonormal basis for the remaining $N-n$ sites of the chain, and  
their internal structure is irrelevant for the remaining discussion. 

The states \eqref{complete} form a complete basis on the 
Wilson chain,\cite{AndersSchiller2005,Anders} and can be used 
to construct the density operator as follows,\cite{Weichselbaum2007} 
\begin{eqnarray}
\varrho &=&\ds \sum_{n=0}^{N}\; \varrho^{[n]}
\;,
\\
\varrho^{[n]} &\equiv& {\sum_{i\in D, e_n}} \sum_{\ug_i}
\frac{e^{-\beta E_i^n}}{\cal Z}
|i, \uG_i,\ug_{i}; e_n\rangle  ^{[n]}\, ^{[n]}
\langle i,\uG_i,\ug_{i}; e_n |\;.\nonumber
%\label{eq:DM}
\end{eqnarray}
Here $\beta =1/k_B T$ and the partition function is expressed as 
\bea
{\cal Z} =\ds{\sum_{n=0}^{N}}\, \ds{\sum_{i\in D}}\, \dim (i)\; e^{-\beta E_i^n}
d^{N-n}\;,
\eea
with $d$ the dimension of the Hilbert space at each added site of the Wilson chain, and $d^{N-n}$
the dimension of the space of the "environment" states, $e_n$. We remark that in the last iteration all states are 
considered to be discarded, while in the first few iterations there are typically no discarded states yet.

To compute local observables and spectral functions of observables at the impurity site,
 one traces out step by step the 
environment states, and introduces  the following set of truncated reduced density matrices, 
\begin{equation}
R^{[n]}\equiv\underset{\left\{ e_n \right\}}{\tr} \left\{\sum_{m \ge n} \varrho^{[m]}   \right\}\; .
\end{equation}
By symmetry, the reduced density matrices are invariant under the symmetries used, and have
therefore a blockdiagonal structure in the representation indices.\cite{Toth}
The matrices $R^{[n-1]}$ can be constructed iteratively from $R^{[n]}$ by tracing out site $n$ and then 
adding the contribution of states  discarded  in iteration $(n-1)\to n$. The contribution of the kept states ($K$) reads:
 \begin{multline}
\langle i\parallel R^{[n-1]} \parallel  j \rangle^{[n-1]}_{i,j\in K} \\
=\ds{\widetilde{ \sum_{{u, v},\mu}}}\;
\frac{\dim(u)}{\dim (i)}\;  \langle u \parallel {R}^{[n]} \parallel  v \rangle^{[n]}\;.
%\delta_{\uG_u, \uG_v}\,.
\end{multline}
Here the tilde indicates that the summation runs over states $u$ and $v$
having the same  symmetry, and 
constructed  from states $i$ and $j$ by adding the \emph{same} local state, 
$i\otimes \mu \to u$ and $j\otimes \mu \to v$.  The subscript indicates that $i$ and $j$ are both \emph{kept states }.
The discarded piece of $R^{[n-1]}$  is then simply
\be
\langle i\parallel R^{[n-1]} \parallel  j \rangle^{[n-1]}_{i,j\in D}
= \delta_{i,j} \frac{d^{N+1-n}\; }{{\cal Z}}\;e^{-\beta E^{n-1}_i}.
\ee

To gain insight to the dynamical properties of a quantum impurity, one usually computes the    retarded Green's functions for some operator multiplets  $A_{\g_A}$ and $B_{\g_B}$,   

\bea
%G^{\rm ret}_{(A_{\g_A})^\dagger,B_{\g_B}}
G^{\rm ret}_{{\g_A},{\g_B}}
(\omega)
%& \equiv & \ds{\int_{-\infty}^{\infty}} G^{\rm ret}_{A,B^\dagger}(t)e^{izt}dt\\
 \equiv  -\frac{i}{\hbar}\ds{\int_{0}^{\infty}}\textrm{Tr}\Bigl\{{\varrho\;}\bigl[({ A}_{\g_A})^\dagger(t),{B}_{\g_B}(0)\bigr]_\xi\Bigr\}\;e^{i\omega t}dt\;,
 \nonumber
 \\
 \label{G} 
 %\nonumber
\eea
with $\xi=-$ ($\xi=+$) referring to commutators (anticommutators) appearing for  bosonic (fermionic) operators. By symmetry, to have a non-vanishing value,
$A$ and $B$ must both transform according the same representation, $\G_A\cong \G_B$, and 
$\g_A\equiv \g_B$ must also be satisfied, $G^{\rm ret}_{\g_A,\g_B} = \delta_{{\g_A},{\g_B}}
G^{\rm ret}_{A^\dagger,B}$. Notice that Eq.~(\ref{G}) is defined in terms of 
$({\hat A}_{\g_A})^\dagger$, transforming according to the conjugate 
representation, $\G_A^{\;\ast}$.\cite{footnote3}  

The expression above can be evaluated in terms of the truncated  density matrices, $R^{[n]}$, and the reduced matrix elements 
$\langle i \parallel A \parallel  j \rangle _{\ua}^{[n]}$ and 
$\langle i \parallel B \parallel  j \rangle _{\ua}^{[n]}$ 
of the operators $A$ and $B$, to   obtain ~\cite{Anders,Toth} 

\begin{widetext}
\begin{multline}
G^{\rm ret}_{A^\dagger,B}(z)=
\ds{\sum^{N}_{n=0}}\;\ds \sum_{i\in D,K}
{\sum_{(j,k)\notin(\rm K,\rm K)}}
\langle i \parallel R^{[n]}\parallel  j\rangle ^{[n]} \times
\left[
\sum_{\ua}
\frac{\langle k \parallel A\parallel  j \rangle _{\ua}^{[n]\ast}
\langle k  \parallel B\parallel  i\rangle_{\ua}^{[n]}}
{z+\frac{1}{2}(E^n_i +E^n_j)-E^n_k}\frac{\dim(k)}{\dim(A)}\right .
\\
\left . -\xi\sum_{\ua}   \frac{\langle j  \parallel B \parallel  k \rangle_{\ua}^{[n]}
\langle i   \parallel A \parallel  k \rangle_{\ua}^{[n]\ast}}
{ z-\frac{1}{2}(E^n_i +E^n_j)+E^n_k } \frac{\dim (i)}{\dim (A)} \right]\;.
\label{eq:green_function_explicite}
\end{multline}
\end{widetext}
This expression provides an efficient way to compute spectral
functions. Notice that it contains only the reduced matrix elements
and the dimensions of the multiplets involved.

\section{Study of the \SU 3 Anderson model}
\label{sec:SU3-Anderson}

To demonstrate how the scheme presented above works, we apply
it to study the repulsive
\SU 3-symmetrical Anderson model, defined already in the
Introduction. We perform our calculations for a conduction
band with a uniform local density of states between energies
$W>\epsilon>-W$ with the bandwidth set to $W\equiv1$, and use the
corresponding hopping amplitudes 
$h^{[n]}\simeq(1/2) (1+\Lambda^{-1})\Lambda^{-n/2} $ 
in \eqref{Wilson_Tau}.
In this case, the width of the (noninteracting) level is approximately
given by
 $\D = \pi \varrho_c \tilde V ^2  $ with
$\varrho_c= 1/2W$ the local density of states at site $0$ of the Wilson
chain.  
\begin{table}[h]
\begin{tabular}{|c|c|c|c|}
\hline
Operator & ($Q, F$) &  Dim & Components \\
\hline\hline
& & & \\
  &     &    &  ${\young (1) } \rightarrow f^{\dagger}_1 $\\
  & & & \\
 $f^{[n]\dagger}_\alpha$, $d^{\dagger}_\alpha$  & $\left( 1, \yng(1)\, \right ) $ & $1\times 3$ & ${\young ( 2)}\rightarrow f^{\dagger}_2$\\
 & & & \\
& & & ${\young ( 3)}\rightarrow f^{\dagger}_3$ \\
& & & \\
\hline
& & & \\
& & & ${\young(1,2)} \rightarrow d^{\dagger}_1 d^{\dagger}_2 $\\
& & & \\
$d_\alpha^{\dagger} d_\beta^{\dagger} $   &   $\left(2, \yng(1,1)\, \right)$  & $ 1\times 3$ &  ${\young 
( 1,3)} \rightarrow d^{\dagger}_1 d^{\dagger}_3 $\\
& & & \\
& & & ${\young(2,3)} \rightarrow d^{\dagger}_2 d^{\dagger}_3 $\\
& & & \\
\hline
& & & \\
& & & ${\young(11,2)} \rightarrow d^{\dagger}_1 d_3 $\\
& & & \\
& & & ${\young (12,2)} \rightarrow d^{\dagger}_2 d_3 $\\
& & & \\
& & & ${\young (11,3)} \rightarrow - d^{\dagger}_1 d_2 $\\
& & & \\
& & & ${\young (12,3)} \rightarrow \frac{1}{\sqrt{2}}\left(d^{\dagger}_1 d_1-d^{\dagger}_2d_2\right) $\\
& & & \\
$d_\alpha^{\dagger} d_\beta$    &   $\left(0, \yng(2,1)\, \right)$  & $1\times 8$ &  ${\young (22,3)} \rightarrow d^{\dagger}_2 d_1 $\\
& & & \\
& & & ${\young (13,2)} \rightarrow \frac{1}{\sqrt{6}}\left(-d^{\dagger}_1 d_1-d^{\dagger}_2d_2 \right. $\\
& & & \\
& & & $\left . +2d^{\dagger}_3d_3\right)$\\
& & & \\
& & & ${\young (13,3)} \rightarrow -d^{\dagger}_3 d_2 $\\
& & & \\
& & & ${\young (23,3) } \rightarrow d^{\dagger}_3 d_1 $\\
& & & \\
\hline
\rule[-.5em]{0pt}{1.7em}
 $d_1^{\dagger}d_2^{\dagger}d_3^{\dagger}$  & $\left( 3, \bullet \right )$ &$ 1\times 1 $& ${ \bullet} \rightarrow d^{\dagger}_1d^{\dagger}_2d^{\dagger}_3$\\
\hline
\end{tabular}
\caption{Irreducible tensor operators for the \SU 3 symmetry.
\label{table:tensors}}
\end{table}

As mentioned before, the Hamiltonians \eqref{eq:H_SUN},
\eqref{Wilson_H0} and \eqref{Wilson_Tau} have a $\text{U}(1)\times \SU 3$
symmetry in the charge and flavor sectors, respectively.
Correspondingly, multiplets of the Hamiltonian 
are characterized by a charge and a flavor quantum
number.  The charge quantum numbers $Q_i$ are simply
identical to the total charge,~\cite{footnote_charge}
\be Q \equiv
\sum_{\alpha}\left \{\sum_{n=0}^{N}\left( f^{[n]\dagger}_{\alpha}
  f^{[n]}_{\alpha}-{3 \over 2}\right) + 
  \left (d_\alpha^{\dagger}\,d_\alpha -{3\over 2}\right ) \right \}, 
\ee 
  conserved by
\eqref{eq:H_SUN}, \eqref{Wilson_H0}, and \eqref{Wilson_Tau}.

Labeling the \SU 3 representations and the states within an
\SU 3 multiplet is somewhat more complicated. The flavor quantum
numbers $F_i$ can be (and are usually) represented by Young tableaux,
characterized by two non-negative integers in case of \SU 3 (see
Appendix~\ref {appendix:SU3}). Young tableaux provide a nice
pictorial way to multiply and decompose representations, or calculate
their dimensions.  For our numerical calculations, however, we had to
construct explicitly the basis states of \SU 3 representations and to
compute the corresponding Clebsch-Gordan coefficients.\cite{footnote4}
This we carried out using the so-called Gelfand-Tsetlin patterns,
briefly discussed in
Appendix~\ref{appendix:SU3}. The states obtained this way are analogues of
the canonical \SU 2 basis states, created by the raising and lowering
spin operators, $S^\pm$.  Gelfand-Tsetlin patterns are in one-to-one
correspondence with the Young tableaux, but they allow for a simpler
explicit construction of the basis states.  For more details, we refer
the reader to Ref.~\onlinecite{ArneJan}.

\begin{figure}
  \begin{center}
  %  \begin{minipage}[t]{0.5\columnwidth}
      \includegraphics[width=0.9\columnwidth, clip]{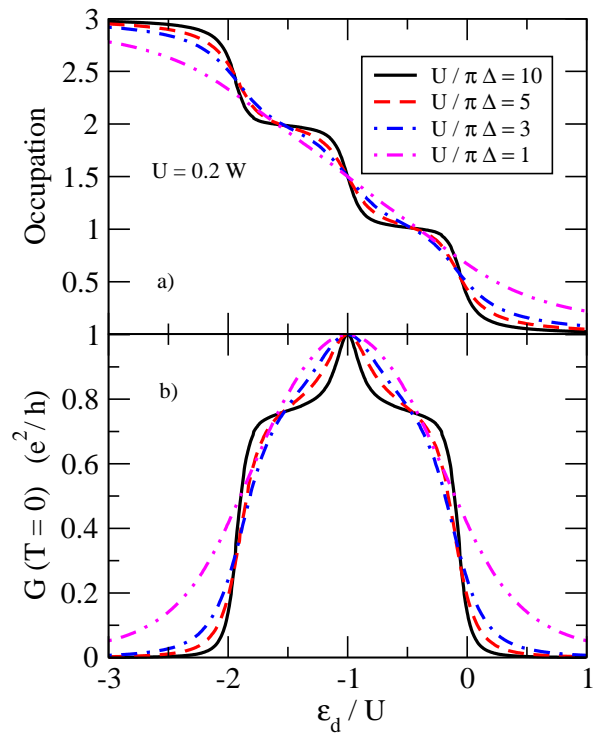}
 %   \end{minipage}\hfill
 %   \begin{minipage}[t]{0.5\columnwidth}
%      \includegraphics[width= 0.98\columnwidth, clip]{conductance_epsilon_d_T_0.eps}
  %  \end{minipage}
   \end{center}
\caption{\label{fig:occupation_conductance_T_0}
    (Color online) The occupation number (upper panel) 
    and the conductance (lower panel) as function of $\varepsilon_d$ and for 
	different broadening parameters $\D$. The temperature is $T=0$ and the
	Coulomb interaction is fixed to $U = 0.2\, W.$
	\label{fig:occupation_conductance}
}
\end{figure}

Similar to the  eigenstates and multiplets, we also need to group operators into \SU 3 multiplets and characterize them by appropriate \SU 3 quantum numbers.
This classification of the most important operators is summarized in Table~\ref{table:tensors}. 
In  terms of \SU 3, there is only a single hopping operator, 
$C^{[n]}_\gamma \leftrightarrow  {f^{[n]\dagger}_\alpha}$, which transforms according to the defining \SU 3 representation and has charge $Q=1$, similar to the creation operator of the localized level, $d^\dagger_\alpha$.  From $d^\dagger_\alpha$ we can also construct various local operators of interest. The  spin operators, $\sim d^\dagger_\alpha d_\beta$ form, e.g., an 8-dimensional charge $Q=0$
operator multiplet  in terms of $\text{U}(1)\times \SU 3$, while the charge $Q=2$ pairing 
operators $\sim d^\dagger_\alpha d^\dagger_\beta$
transform according to a 3-dimensional \SU 3 representation. Finally, the "trion" 
operator $d^\dagger_1d^\dagger_2d^\dagger_2$ has charge $Q=3$ and is an \SU 3 singlet.  

\subsection{Numerical results}

In our runs we have kept about 250 multiplets at each 
iteration, corresponding to approximately 1500 states on average, while 
the Wilson parameter was fixed to $\Lambda =2$.

\begin{figure}[t]
  \includegraphics[width=0.98\columnwidth]{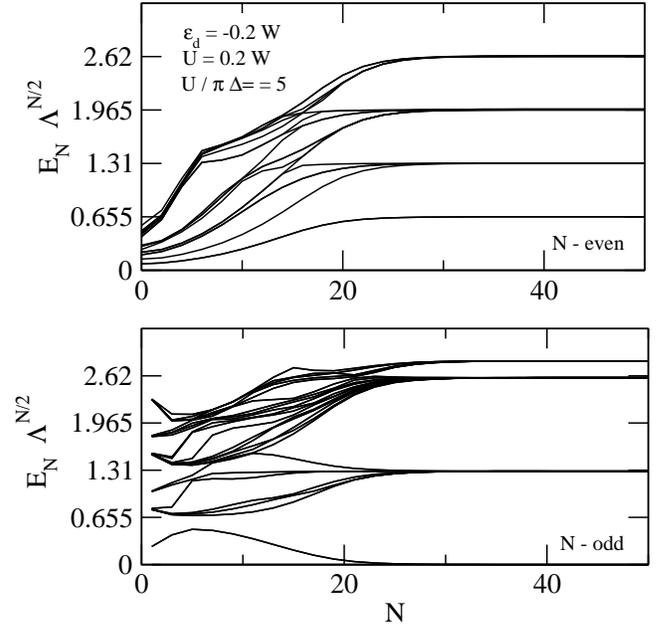}
  \caption{\label{fig:flow_diagram_half_filling}
 NRG finite size spectrum at half-filling and T=0. 
 Upper (lower) panel represents the even
 (odd) part of the spectrum. In both plots we represent the lowest 50 energy levels. The parameters are: $\Lambda = 2$, $U = 0.2 \;W$, $\e_d = -U$ and 
 $U/\pi\D = 5$. The convergence was reached after approximately 27 iterations. 
 \label{fig:flow_diagram} }
\end{figure}

In Fig. \ref{fig:occupation_conductance}(a)  we
present the occupation of the localized level as a function of the
energy $\e_d$.  The localized level can accommodate up to 3 fermions.
For large enough $U/\Delta$'s, these fermions enter the local level one
by one as $\e_d$ is decreased, and a
Coulomb staircase is clearly visible. In the range $-U<\e_d<0$ there
is approximately one electron on the level. The \SU 3 spin of this
electron is screened by the conduction electrons, and an \SU 3 Kondo
state is formed. A similar, hole-like $\SU 3^*$ state emerges for
$-2U<\e_d<-U$.  

\begin{table}[t]
Even iterations\vspace{0.3cm}
\begin{tabular}{|c|c|c|c|c|}
\hline
$E/({2\pi/L})$ & $E_{NRG}/ E_L$ &  $Q$ (charge) & \SU 3 & dimension\\
\hline\hline
0    &   0   &   $Q_0$ &  \textbullet &  1\\
\hline
& & & &  \\
$1/2$ & $0.5$ &  $Q_0+1$ &$ \yng (1)$ & 3\\
& & & &  \\
$1/2$ & $0.5$ & $Q_0-1$ &$ \yng (1,1)$ & 3\\
& & & &  \\
\hline
& & & &  \\
$1 $ & $1$ &  $Q_0+2$ &$ \yng(1,1)$ & 3\\
& & & &  \\
$1$ &  $1$ & $Q_0$  &$ \yng(2,1)$ & 8\\
& & & &  \\
$1$ & $1$ & $Q_0 $ & \textbullet & 1\\
& & & &  \\
$1$ & $1$ & $Q_0-2$ &$ \yng(1)$ & 3\\
& & & &  \\
\hline

\hline
\end{tabular}\\
\vspace{0.3cm}
Odd iterations\vspace{0.3cm}
\begin{tabular}{|c|c|c|c|c|}
\hline
$E/({2\pi/L})$ & $E_{NRG}/E_L$ &  $Q$ (charge) & \SU 3 & dimension\\
\hline\hline
0    &   0   &   $Q_0$ &  \textbullet &  1\\
0    &   0   &   $Q_0$ - 1&  $\yng(1,1) $ &  3\\
& & & &  \\
0    &   0   &   $Q_0$ - 2&  $\yng(1) $ &  3\\
0    &   0   &   $Q_0$ - 3&  \textbullet &  3\\
\hline
& & & &  \\
$1 $ & 0.98 & $Q_0$ + 1&$ \yng(1,1)$ & 3\\
& & & &  \\
$1$ & 0.98 & $Q_0$  &$ \yng(2,1) $ & 8\\
$1$ & 0.98 & $Q_0$  & \textbullet & 1\\
$1$ & 0.98 & $Q_0$ - 1 &$ \yng(1,1)$ & 3\\
& & & &  \\
$1$ & 0.98 & $Q_0$ - 1 &$ \yng(2)$ & 6\\
& & & &  \\
$1$ & 0.98 & $Q_0$ - 1 &$ \yng(1,1)$ & 3\\
& & & &  \\
$1$ & 0.98 & $Q_0$ + 1 &$ \yng(2,2)$ & 6\\
& & & &  \\
$1$ & 0.98 & $Q_0$ + 2 &$ \yng(1)$ & 3\\
& & & &  \\
\hline
\end{tabular}
\caption{ Energy spectrum for the electron-hole symmetric, mixed valence
point. $E_L$ stands for the finite size energy scale, $E_L = 2\pi/L = 1.311$.
$Q_0 =0\, (3/2) $ for even (odd) iterations.
\label{table:spectrum_mixed_valence}}
\end{table}

The most intriguing region is though the mixed valence
region, $\e_d\approx -U$. For $\e_d=-U$ the ground state of the
isolated impurity ($\Gamma=0$) would be sixfold degenerate due to
electron-hole symmetry,  connecting the two \SU 3 triplets, 
$\{ d^\dagger_\alpha|0\rangle\}$ and $\{ d^\dagger_\alpha d^\dagger_\beta |0\rangle\}$. 
 Valence fluctuations produce a state with
all these states strongly mixed by quantum fluctuations.  
Fig.~\ref{fig:flow_diagram} and Table \ref{table:spectrum_mixed_valence} show the 
flow diagram of the NRG levels, and the \SU 3 classification of the asymptotic  
finite size spectrum, respectively. A detailed analysis reveals that this finite size spectrum 
can simply be understood as the finite size spectrum of three 
chiral fermions with a phase shift $\delta=\pi/2$ at the Fermi energy.  
This phase shift is indeed in  full agreement with the Friedel
sum rule, $3\delta/\pi =\langle n\rangle $, and the occupation $\langle n\rangle=3/2$ assured by
electron-hole symmetry.

Similarly, the \SU 3 Kondo spectrum, displayed in Table \ref{table:spectrum_kondo}
can be understood as the finite size spectrum of three 
chiral fermions with a phase shift $\delta\approx\pi/3$, implied by the Friedel sum rule 
and the occupation $\langle n\rangle\approx 1$.

\begin{table}[t]
Even iterations\vspace{0.3cm}
\begin{tabular}{|c|c|c|c|c|}
\hline
$E/({2\pi/L})$ & $E_{NRG}/E_L$ &  $Q$ (charge) & \SU 3 & dimension\\
\hline\hline
0    &   0   &   $Q_0$ &  \textbullet &  1\\
\hline
& & & &  \\
$2/3 $ & 0.33 & $Q_0$ - 1 &$ \yng(1,1)$ & 3\\
& & & &  \\
\hline
& & & &  \\
$4/3$ & 0.66 & $Q_0$ + 1 &$ \yng(1)$ & 3\\
& & & &  \\
$4/3$ & 0.66 & $Q_0$ - 2 &$ \yng(1)$ & 3\\
& & & &  \\
\hline
& & & &  \\
$2 $ & 1 & $Q_0$  &$ \yng(2,1)$ & 8\\
$2 $ & 1 & $Q_0$  & \textbullet & 1\\
$2 $ & 1 & $Q_0$ - 3 & \textbullet & 1\\
& & & &  \\
\hline
\end{tabular}\\
\vspace{0.3cm}
%\end{table}
Odd iterations\vspace{0.3cm}
%\begin{table}[t]
\begin{tabular}{|c|c|c|c|c|}
\hline
$E/({2\pi/L})$ & $E_{NRG}/E_L$ &  $Q$ (charge) & \SU 3 & dimension\\
\hline
0    &   0   &   $Q_0$ &  \textbullet &  1\\
\hline 
& & & &  \\
$ 1/3$ & $0.15$ &  $Q_0+1$  & $ \yng(1) $ & 3\\
& & & &  \\
\hline
& & & &  \\
$2/3$ & 0.32 & $Q_0 + 2$ & $ \yng(1,1)$ & 3\\
& & & &  \\
\hline
\end{tabular}
\caption{ Energy spectrum in the Kondo regime. $Q_0 = 0\,(-3/2)$ for 
even (odd) iterations. 
\label{table:spectrum_kondo}}
\end{table}

\begin{figure}[t]
\includegraphics[width=0.93\columnwidth, clip]{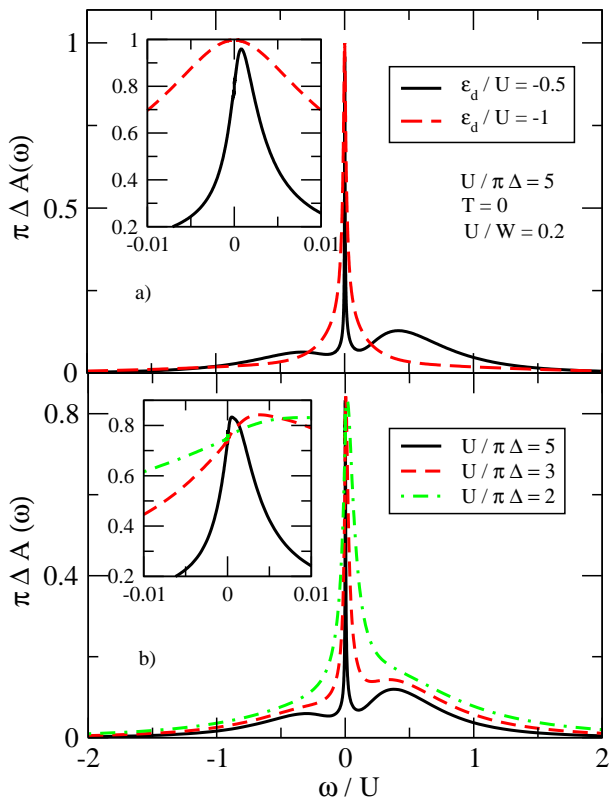}
%{spectral_function_d_T_0_U_over_Gamma_5.eps}
%\includegraphics[width= 0.93\columnwidth, clip]{spectral_function_d_T_0.eps}
\caption{\label{fig:spectral_function_f_dagger} 
    (Color online) a) The normalized spectral function for the 
    on-site creation operator $d^\dagger_\alpha$ for two different fillings $n = 1.5$ 
    (half filling, red dashed) and $n=1$ (1/3 filling, black solid).  In the singly occupied case, $n=1$, 
    the dot is in the Kondo regime, and the resonance is shifted away from the Fermi 
    energy, $\omega=0$. The value of the spectral function at the Fermi energy is determined by the Friedel sum rule, $\pi \D A(0)\to 3/4$. 
    b) Evolution of the normalized spectral function 
    as function of $U/\pi\Delta$ 
    in the Kondo regime, for    $\varepsilon_d /U = -0.5$. The deviation from the Friedel sum rule is 
    less than 1\%.}
\end{figure}

The crossover between the two \SU 3 Kondo regimes through the mixed valence regime is 
maybe best captured by the local level's spectral function, shown in Fig.~\ref{fig:spectral_function_f_dagger}. In the \SU 3 and $\SU 3^*$ Kondo regimes we find 
a  narrow Kondo resonance of an exponentially small width pinned somewhat 
asymmetrically to the Fermi energy, $\omega=0$ (see inset), and two Hubbard peaks. 
At the mixed valence point, $\e_d=-U$
on the other hand, a relatively broad and symmetrical resonance of width $\sim \Gamma$ appears
at the Fermi energy, and the charging peaks are absent. 

\begin{figure}[t]
    \begin{center}
       \includegraphics[width=0.6\columnwidth, clip]{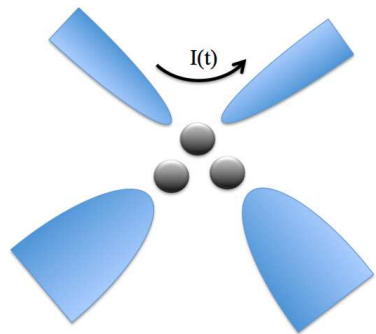}
    \end{center}
\caption{
\label{fig:sketch}
(Color online)  Possible realization of \SU 3 Kondo states using four 
    edge states and three quantum dots. The upper edge state is splitted and 
    allows transport measurements between the two upper external leads.
    \cite{Carmi} }
\end{figure}

\subsection{Conductance}

Let us now consider the mesoscopic structure proposed in Ref.~\onlinecite{Carmi}, sketched in 
Fig.~\ref{fig:sketch}.   As explained in Ref.~\onlinecite{Carmi}, this quantum dot structure would 
possibly realize a Kondo state with an approximate \SU 3-symmetry. 
The conductance through the device  can be directly related to the spectral functions of the 
$d$-level,  and for a symmetrical device one finds \be
G(T) = \pi \Delta \;{e^2 \over h } \int_{-\infty}^{\infty}{\rm d}\omega\;
A(\omega, T) \left (-\frac{\partial f(\omega)}{\partial \omega} \right)\;.
\label{eq:conductance}
\ee

Here $A(\omega)$ is the spectral function of the $d_\alpha$ operator of the localized
level, $ A(\omega) = -\Im m\, G_{d_\alpha\,d_\beta^\dagger} (\omega)/\pi $, with 
$G_{d_\alpha\, d_\beta^\dagger} (\omega)$ the Green's function defined in Eq.~\eqref{G}. 
The corresponding $T=0$ temperature linear conductance is shown in 
Fig.~\ref{fig:occupation_conductance}(b)
 as a function of the energy $\epsilon_d$. 
The conductance reaches its maximal value in the mixed valence regime, 
$\e_d\sim -U$, and displays \SU 3 Kondo effect related plateaus in the regions
$-2U<\e_d < -U$  and $-U<\e_d < 0$. The $T=0$ temperature conductance values observed  
can be understood in terms of the Friedel sum rule, relating the total occupation of the $d$-level to the phase shift of the conduction electrons,
which yields $3 \delta/\pi = \langle n\rangle$ for the
\SU 3 Anderson model.~\cite{Carmi}
In a Fermi liquid state, --- at $T=0$ temperature, ---  the conductance can be computed using the Landauer-B\"uttiker formula, and in the geometry 
of Fig.~\ref{fig:sketch}
is simply related to the phase shift as $G(T=0)=(e^2/h) \sin^2(\delta)$. This explains the value 
$G_{\SU 3}\approx (3/4)\; e^2/h$ observed in the Kondo states; there the occupancies 
are  $\langle n\rangle\approx 1$ and $\langle n\rangle\approx 2$, corresponding to phase shifts 
$\delta=\pm \pi /3$, and the previously mentioned value of the conductance.  
 At the mixed valence point, 
$\e_d=-U$, on the other  hand, we have  $\langle n\rangle = 3/2$, implying a phase shift $\delta=\pi/2$, and a maximal conductance, $G_{\SU 3} = e^2/h$ .

\begin{figure}[b]
\begin{center}
\includegraphics[width=1\columnwidth, clip]{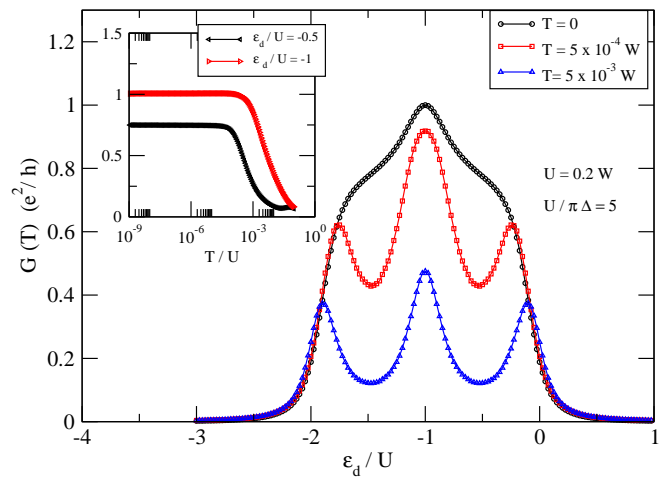}
\end{center}
\caption{   \label{fig:conductance_finite_T}
 (Color online)  Finite temperature conductance as function of 
    $\varepsilon_d$. Inset: The temperature dependence of the conductance
    on a logarithmic scale,	for two different fillings, 
    $n=1.5$ (half filling, red dashed) and $n=1$ (1/3 filling, black solid).
    In both panels the Coulomb energy and the broadening to the 
    leads were fixed to $U = 0.2\,W$ and $U/\pi\D = 5$.}
\end{figure}

\begin{figure}
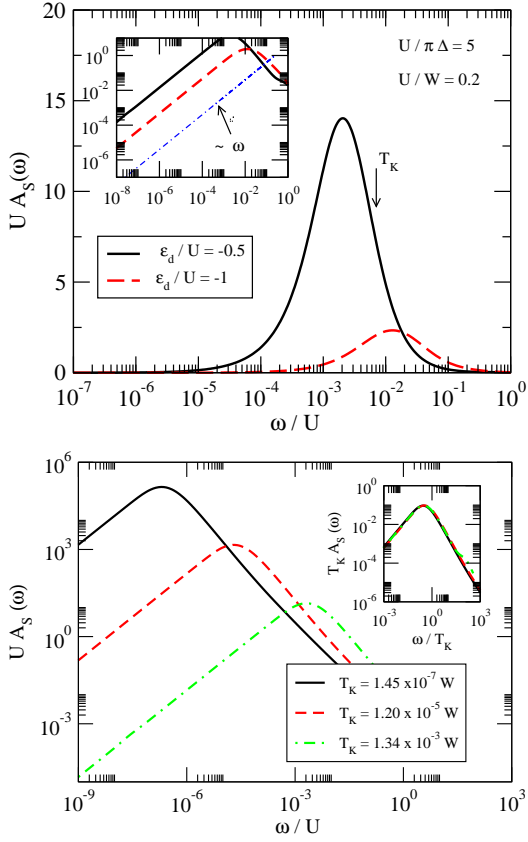

      \includegraphics[width=0.8\columnwidth, clip]{Fig6a.eps}
      \vskip0.2cm
      \includegraphics[width= 0.8\columnwidth, clip]{Fig6b.eps} 
       \caption{   
      \label{fig:spin}
     (Color online) Upper panel: Spectral function of the spin operator $d^\dagger_\alpha d_\beta$ on a 
    logarithmic scale for two different fillings, $\langle n\rangle= 1.5$ 
    (half filling, red dashed) and $\langle n\rangle \approx 1$ (1/3 filling, black solid) for  
    $U = 0.2\,W$.
    In the singly occupied case, $\langle n\rangle \approx 1$,  the dot is in the Kondo regime. The inset indicates the linear decay 
    in $A_{S}(\omega)$ in the small frequency limit. 
    The blue, dashed-dotted line is a guideline for the eye.  
    Lower panel: The spectral function of the spin operator
    in the Kondo regime, $\langle n\rangle \approx 1$, and for three different ratios $U/\pi\D$: $U/\pi\D=15$ (solid black line), $U/\pi\D=10$ (dashed red line) and
      $U/\pi\D=5$ (dashed-dotted green line). The inset indicates the
    universal scaling  collapse of the spin spectral function in 
    the Kondo regime. 
      }

\end{figure}

Increasing the temperature, the conductance is quickly suppressed in
the Kondo regimes, and three Coulomb blockade conductance peaks emerge
at the points of charge degeneracy, as shown in
Fig.~\ref{fig:conductance_finite_T}.  The central peak corresponds to
the transition $n=1\leftrightarrow 2$ while the two side peaks
correspond to charge fluctuations $n=1\leftrightarrow 0$ and
$n=2\leftrightarrow 3$, respectively.  In
Fig.~\ref{fig:conductance_finite_T} we also display\scrap{ed} the
temperature dependence of the conductance at the mixed valence point
and in the \SU 3 Kondo regime. The Kondo temperature of the \SU 3
Kondo state is clearly much smaller than the mixed valence energy
scale even for these moderate interactions.  This Kondo temperature can be estimated 
by first doing perturbation theory in $\tilde V$ and performing a Schrieffer-Wolff transformation and 
then carrying out a renormalization group analysis. This analysis yields a Kondo temperature
of
\be
T^{\SU 3}_K \approx D_0 \; e^{-1/3 \lambda} \;, 
\label{eq:TK}
\ee
with the dimensionless coupling $\lambda$ expressed as 
\be
\lambda = \frac{\Delta}{\pi E_+}  + \frac{\Delta}{\pi  E_-}\;
\ee
in terms of the level width $\Delta$ and the  "ionization energies"
$E_+ = U+\epsilon_d$ and $ E_- =- \epsilon_d$, and 
$D_0\approx \min (E_+, E_-)$ a high energy cut-off.

\begin{figure}[b]
     \includegraphics[width= 0.8\columnwidth, clip]{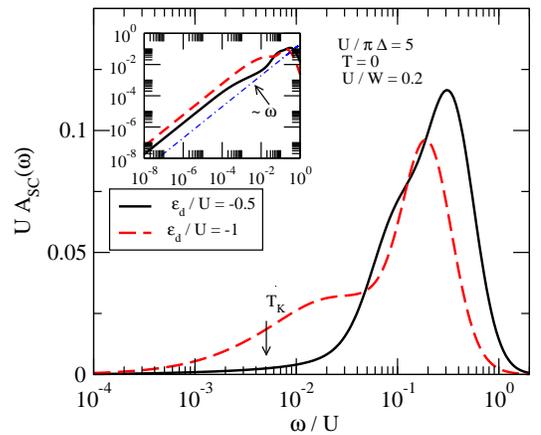}
\caption{
      (Color online) 
    Spectral function of the superconducting operator for  
    $U = 0.2\,D$ and $\langle n\rangle= 1.5$ 
    (half filling, red dashed) and $\langle n\rangle \approx 1$ (1/3 filling, black solid).
          \label{fig:SC} 
    }
\end{figure}

\subsection{Spectral functions of local operators}

The behavior of strongly correlated 
cold atomic and heavy fermion lattice systems can often be understood in terms of a self-consistent
quantum impurity model (dynamical mean field theory). Within this picture,  the local response functions of the quantum impurities  may drive superconducting or magnetic phase transitions, or lead to even more exotic quantum phases. 
In this subsection, let us therefore analyze the spectral properties 
of the \SU 3 Anderson model, and investigate 
  the local spectral and response functions
  of  its spin, pairing, and trion operators.

\begin{figure}[b]
     
      \includegraphics[width= 0.9\columnwidth, clip]{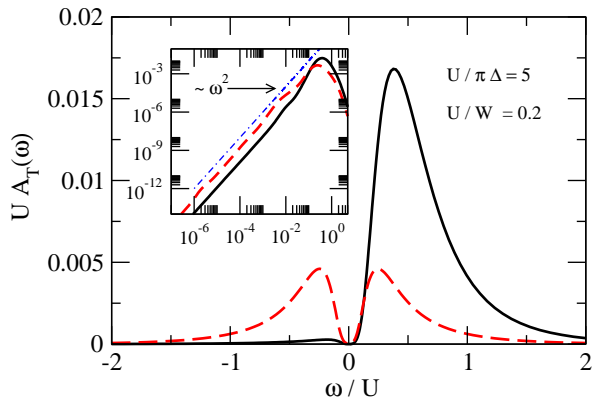}
\caption{
     The spectral function for the trion operator  
    for two different fillings $n = 1.5$ 
    (half filling, red dashed) and $n=1$ (1/3 filling, black solid).
    The inset represents the same data on a 
    logarithmic scale. The trionic spectral function
    scales as $\omega^2$ in the small energy limit.     
    \label{fig:trion} 
    }
\end{figure}

As discussed before, the spin operators, $d^\dagger_\alpha d_\beta$ transform according to an 8-dimensional \SU 3 representation. As shown in  Fig.~\ref{fig:spin}, 
their spectral function displays Fermi liquid properties (see inset), and behaves very similarly to the spin spectral function of a standard \SU 2
Anderson model.\cite{Hewson2006} In the mixed valence regime, for $\D\ll U$ 
charge fluctuations to the state $n=0$ and $n=3$ are frozen out, and at low energies  
the only relevant energy scale is  $\D$; correspondingly, the spectral function exhibits a broad resonance at $\omega\sim \D$ (extending up to $\omega\sim U$), and decays linearly to zero for small frequencies, $A_S^{n\approx 3/2}(\omega)\sim \omega/\D^2$. 
By Hilbert transform, this amounts in a spin 
susceptibility, $\chi_S\sim 1/\D$. In the Kondo regime, $\langle n\rangle\approx 1$, on the other hand, 
two separate scales can be distinguished. Below $\omega\sim \min(U,|\epsilon_d|)$ charge fluctuations are frozen and a clear Kondo anomaly can be observed as a logarithmic increase of the spectral 
function, as indicated by the arrow in Fig.~\ref{fig:spin}.
The Fermi liquid behavior only emerges below  the Kondo scale, $\omega<T_K\ll \D,U$. 
In this Kondo regime $A_S^{\rm Kondo}(\omega)\sim \omega/T_K^2$, and correspondingly, 
a susceptibility $\chi_S\sim 1/T_K$ is found. 

The spin spectral function becomes universal in the Kondo limit, $T_K\ll \Delta,U$ in the sense that 
the $T=0$ temperature dynamical susceptibility scales as 
 $\chi_S(\omega)=  (1/T_K)\; f(\omega/T_K,\langle n\rangle)$, with $f(\omega/T_K,\langle n\rangle)$
a function, which only slightly depends on the occupation of the level,  $\langle n\rangle\approx 1$. 
This is demonstrated in the lower panel of Fig.~\ref{fig:spin} for the imaginary part of the susceptibility, 
$\chi^{\prime\prime}_S(\omega) =  - \pi A_S(\omega)$, computed for different values of $U/\pi\D$.
Numerically we define $T_K$ as the half width at half maximum of $A(\omega)$, the
spectral function of the $d$ operator of the localized level.

%We checked that this is consistent, up to a prefactor of $\simeq 2$,
%with the formula given in Eq. \eqref{eq:TK}.

The correlations of the pairing operators,  $d^\dagger_\alpha d^\dagger_\beta$, behave somewhat similar to those of the spin in the sense that at small frequencies a linear frequency dependence is found, corresponding to a Fermi liquid state with a constant pairings susceptibility (see inset of Fig.~\ref{fig:SC}). However, the local pairing operators  lead out of the local low energy 
charging states both in the mixed valence and in the Kondo regimes. Therefore, the 
amplitudes of their spectral functions as well as the 
corresponding pairing susceptibilities 
obtained by a Hilbert transform,
are typically small, and they are not expected to drive any transition.

As a final example, we display the spectral function of the \SU 3
singlet trion operator, $T^\dagger =
d^\dagger_1d^\dagger_2d^\dagger_3$ in Fig.~\ref{fig:trion}. The trion
operator plays an important role in the attractive
case,\cite{Klingschat2010} however, in this repulsive model it
is a highly suppressed operator. In the Kondo regimes, it has a
non-zero spectral function only because the Kondo states have a small
 ($\sim \D/U$) admixture of the empty and the triply occupied
states, respectively. This explains why the amplitude of the signal is
in this case smaller in the mixed valence regime. It also explains the
strong electron-hole asymmetry in the Kondo regimes. In the $\langle
n\rangle \approx 1$ regime, shown in Fig.~\ref{fig:trion}, e.g., the
admixture of the $n=0$ state is relatively large, $\sim \D/U$,
while the $n=3$ state has a much smaller weight, $\sim
(\D/U)^2$. As a consequence, most of the spectral weight appears
on the particle-like side of the spectral function, $\omega>0$. At
small frequencies the spectral function decays as $\sim
\omega^2$. This is in agreement with Fermi liquid theory, which would
predict a $\langle T(t) T^\dagger(0)\rangle \sim 1/t^3 $ decay of the
trionic correlation functions at very long times.

\section{Conclusions}
\label{sec:conclusions}

In this paper, we showed how to extend the DM-NRG scheme of
Ref.~\onlinecite{Toth} to symmetries with outer multiplicities. As an
application, we performed a detailed DM-NRG study of the \SU 3
symmetrical Anderson model by first incorporating \SU N
symmetries\cite{ArneJan} in the Open Access Budapest DM-NRG
code,\cite{OpenAccess} and then performing the numerical calculations
using the complete $\text U(1)\times \SU 3$ symmetry of the model.  A
similar extension has been carried out within the matrix product state
(MPS) approach parallel to this work.\cite{Andreas}

The properties of the \SU 3 Anderson model do not differ so much from those of the original Anderson model. As also discussed in  Refs.~\onlinecite{Carmi}, for $U>\D$
four distinct charging regions appear: the featureless empty and 
fully occupied  regions ($\langle n\rangle \approx 3$ and $\langle n\rangle \approx 0$), and 
two Kondo regions of occupancies $\langle n\rangle \approx 1$
and $\langle n\rangle \approx 2$, respectively. 
The two \SU 3 Kondo regions  behave similarly: they are characterized by phase shifts 
$\delta \sim \pm \pi/3$, as verified from the finite size spectrum, and correspondingly, a Kondo resonance shifted away from the Fermi energy.  
In these Kondo regimes, the susceptibility has a universal
 form, $\chi(\omega)= f(\omega/T_K, \langle n\rangle) / T_K$, with a scaling function $f(x,\langle n\rangle)$ very similar to the one emerging in the \SU 2 
Anderson and Kondo  models. 
For completeness, we also studied the spectral properties of other local operators such as pairing or  the trion operators. Both of them turn out to have a  small amplitude for $\D\ll U$, and exhibit simple Fermi liquid properties below the mixed valence  and  Kondo scales, respectively. 
Therefore, away from half filling,  a magnetic instability is expected to prevail on a lattice
in the \SU 3 Hubbard model, in general agreement with the results of Gutzwiller calculations
at low temperatures.\cite{RappRosch}

The \SU 3 Kondo regions are separated by a mixed valence state, which again has a Fermi liquid 
character with a Fermi liquid scale  of the order of the level width, $\G$. Here we find a phase shift 
$\delta = \pi/2$, in agreement with the expectations based upon the Friedel sum rule, but apart from that, and the emerging electron-hole symmetry at this point, the properties of the mixed valence state appear to be very quite similar to those of the Kondo states.

We also investigated the conductance properties of the \SU 3
arrangement, proposed in Ref.~\onlinecite{Carmi}. At high temperatures
we observe in the side-conductance three charging peaks, corresponding
to the three charging steps. As the temperature is lowered, the
Coulomb blockade valleys are gradually filled up, and two conductance
shoulders emerge in the Kondo regime with a conductance $G\approx
(3/4)(e^2/h)$.

The methods and the computations presented here represent a first and
important step to perform DM-NRG and DMFT calculations for more
elaborate \SU N or $\text{Sp}(N)$ lattice models.  However, further work is
necessary to optimize
these DMFT calculations. While we definitely gain
enormous storage space by using \SU N symmetries, the evaluation of
the reduced matrix elements and multiple sums over internal
representation labels are currently not carried out with maximal efficiency.
 Since for large $N$'s the dimensions of irreducible \SU N
representations grow very fast, these summations quickly become the
bottleneck for DM-NRG calculations, and further work is needed to
increase the efficiency of the corresponding subroutines.

\emph{Acknowledgment:} 
We acknowledge numerous fruitful discussions with A. Weichselbaum.
We thank the National Institute for Theoretical Physics in Stellenbosch,
South Africa, where this collaboration was started, for hospitality,
and the DFG (De730/7-1) for financially supporting our visit there.
This research has been supported by Hungarian research funds
OTKA and NKTH under Grant Nos.~K73361 and CNK80991, 
and the EU-NKTH GEOMDISS project. 
JvD and AA acknowledge financial support from SFB-TR12. 
CPM has benefit from financial support from 
UEFISCDI under French-Romanian Grant 
DYMESYS (ANR 2011-IS04-001-01 and Contract
No. PN-II-ID-JRP-2011-1).

\appendix

\section{Some details on \SU N representations}
\label{appendix:SU3}
%\begin{itemize}
%\item generators
%\item raising lowering ops
%\item Cartan subalgebra
%\item weights
%\item inner multiplicity
%\item irrep labeling
%\item dimension
%\item product decomposition
%\item Littlewood-Richardson
%\item outer multiplicity
%\item reference to matrix elements
%\item gt patterns to young tableaux
%\item highest-weight state
%\item selection rules for CGCs
%\item generalization of Wigner-Eckart theorem
%\item irreducible tensor ops?
%\end{itemize}

In this appendix we give a brief overview of the representation theory of \SU N,
following the approach of Ref.~\onlinecite{ArneJan},
where a more detailed discussion can be found.

While \SU 2 has three generators, $\mathcal{\hat J}_z$, $\mathcal{\hat J}_+$, and $\mathcal{\hat J}_-$,
\SU N has $N^2-1$ generators.
We shall deal explicitly with only $3(N-1)$ of them,
denoted by $\mathcal{\hat J}^{(l)}_z$ and $\mathcal{\hat J}^{(l)}_\pm$,
where $l=1,\dotsc,N-1$.
By definition, they satisfy the commutation relations
\begin{equation}
\label{eq:su3-commutators}
\left[ \mathcal{\hat J}_z^{(l)}, \mathcal{\hat J}_\pm^{(l)} \right] =  \pm \mathcal{\hat J}_\pm^{(l)}, \quad
\left[ \mathcal{\hat J}_+^{(l)}, \mathcal{\hat J}_-^{(l)} \right] = 2 \mathcal{\hat J}_z^{(l)}.
\end{equation}
These have the same form as those of the corresponding $N\times N$ 
 matrices
\begin{subequations}
\label{eq:defining_representation}
\begin{align}
J^{(l)}_z &= \kbordermatrix{ &&&& \substack{l \\ \downarrow} & \substack{l+1 \\ \downarrow} \\ & 0 \\ && \ddots \\ &&& 0 \\ l\rightarrow &&&& \frac12 & 0 \\ l+1\rightarrow &&&& 0 & -\frac12 \\ &&&&&& 0 \\ &&&&&&& \ddots \\ &&&&&&&& 0 }, \\
J^{(l)}_+ &= \kbordermatrix{ &&&& \substack{l \\ \downarrow} & \substack{l+1 \\ \downarrow} \\ & 0 \\ && \ddots \\ &&& 0 \\ l\rightarrow &&&& 0 & 1 \\ l+1\rightarrow &&&& 0 & 0 \\ &&&&&& 0 \\ &&&&&&& \ddots \\ &&&&&&&& 0 }, \\
J^{(l)}_- &= \kbordermatrix{ &&&& \substack{l \\ \downarrow} & \substack{l+1 \\ \downarrow} \\ & 0 \\ && \ddots \\ &&& 0 \\ l\rightarrow &&&& 0 & 0 \\ l+1\rightarrow &&&& 1 & 0 \\ &&&&&& 0 \\ &&&&&&& \ddots \\ &&&&&&&& 0},
\end{align}
\end{subequations}
which generate the \textit{defining representation} of \SU N.
The $N^2-3N+2$ remaining generators are obtained as commutators
between generators involving different values of $l$.
Below we will use 
$\mathcal{\hat J}^a$ (and $J^a$) as collective notation 
for any of the generators (and corresponding matrices in
the defining representation).

Let $\ket{\Gamma,\gamma}$ denote the
basis states of a general \SU N irrep, where $\Gamma$ labels the
irrep  and $\gamma$ its individual basis states.
%\arne{To describe the transformation properties of a basis state
%  under \SU N rotations, it needs a label $\Gamma$, designating an \SU
%  N irrep, and a label $\gamma$, designating a particular state
%  $\ket{\Gamma,\gamma}$ inside of $\Gamma$.  
%Since the irrep is invariant under the action of the group, 
The action of any  generator on a basis state can be
written as
\begin{equation}
\mathcal{\hat J}^a \ket{\Gamma,\gamma} = \sum_{\gamma'} \Gamma^a_{\gamma',\gamma} \ket{\Gamma,\gamma'},
\label{eq:multiplet_transformation}
\end{equation}
where $\Gamma^a_{\gamma',\gamma} \equiv \braket{\Gamma,\gamma' |
  \mathcal{\hat J}^a | \Gamma,\gamma}$ are the matrix elements of
$\mathcal{\hat J}^a$ within the irrep $\Gamma$.  For example, the
matrix elements of $\mathcal{\hat J}^{(l)}_{z,\pm} $ in the defining
representation are given by Eqs.~(\ref{eq:defining_representation}).

It is convenient to choose the states $|\Gamma,\gamma \rangle$
%It is desirable to choose eigenstates of a Hamiltonian with \SU N
%symmetry 
to be simultaneous eigenstates of $\mathcal{\hat J}^{(l)}_z$
for $l=1,\dotsc,N-1$, with eigenvalues $\lambda^{\Gamma,\gamma}_l$, say.
The sequence
$W_z(\Gamma,\gamma)=(\lambda^{\Gamma,\gamma}_1,\dotsc,\lambda^{\Gamma,\gamma}_{N-1})$
is called its \emph{weight}.  As for \SU 2, all components of the weight takes half-
integer values.

\begin{figure}
\includegraphics[width=\columnwidth, clip]{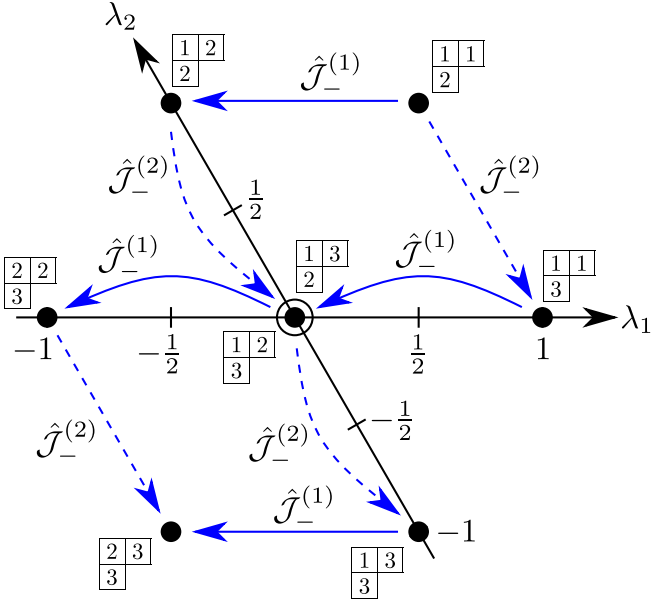}
\caption{Weight diagram of the \SU 3 irrep $\tiny\protect\yng(2,1)$
(non-orthogonal axes chosen to emphasize the symmetric structure of the irrep).
Each dot represents a weight; we also indicate the Young tableaux of the corresponding states.
The circled dot indicates a weight with inner multiplicity $2$.
The blue solid and dashed arrows represent the action of $J^{(1)}_-$ and $J^{(2)}_-$, respectively.}
\label{fig:weight-diagram}
\end{figure}

A convenient way to visualize all states of the same \SU N irrep
is then given by  \emph{weight diagrams} (see Fig.~\ref{fig:weight-diagram} for an example),
which are constructed by marking the point with coordinates $W_z(\Gamma,\gamma)$
for each state $\ket{\Gamma,\gamma}$ of an irrep $\Gamma$.
The operators $\hat J^{(l)}_\pm$ then map states 
onto their neighbors in a weight diagram
($\lambda^{\Gamma,\gamma}_l \to \lambda^{\Gamma,\gamma}_l \pm 1$).

Each irrep $\Gamma$ has a so-called \emph{highest-weight state} (unique
up to a phase), denoted by $\ket{\Gamma,\gamma=\Gamma}$ for convenience.
It is annihilated by all $\mathcal{\hat J}^{(l)}_+$,
\begin{equation}
\mathcal{\hat J}^{(l)}_+ \ket{\Gamma,\gamma=\Gamma} = 0 \quad (l=1,\dotsc,N-1).
\end{equation}
Its weight actually determines the properties of the whole irrep $\Gamma$,
and is thus suitable to provide a labeling scheme for $\Gamma$.

In contrast to \SU 2, several states of an irrep $\Gamma$
can have the same weight. The number of states with the same weight
is called the \emph{inner multiplicity} of this weight.
Consequently, weights are not suitable as the label $\gamma$.
Instead, we use one of two equivalent labeling schemes,
\emph{Young tableaux} or \emph{Gelfand-Tsetlin patterns}.

An \SU N Young tableau is a single, contiguous cluster of left-aligned boxes
with at most $N$ rows, such that each row is not longer than the one above. 
Each box of a tableau carries a number between $1$ and $N$, inclusive,
such that numbers do not decrease from left to right,
and numbers increase strictly from top to bottom.

A \GTP{} $M$ is a triangular matrix of integers $M_{k,l} \; (1 \le k \le l \le N)$,
commonly written as
\begin{equation}
\label{eq:gtpattern}
M = \begin{pmatrix}
\multicolumn{2}{c}{m_{1,N}} & \multicolumn{2}{c}{m_{2,N}} &
\multicolumn{2}{c}{\ldots} & \multicolumn{2}{c}{m_{N,N}} \\
& \multicolumn{2}{c}{m_{1,N-1}} & \multicolumn{2}{c}{\ldots}
& \multicolumn{2}{c}{m_{N-1,N-1}} & \\
&& \ddots &&& \reflectbox{\(\ddots\)} && \\
&& \multicolumn{2}{r}{m_{1,2}} & \multicolumn{2}{l}{m_{2,2}} && \\
&&& \multicolumn{2}{c}{m_{1,1}} &&&
\end{pmatrix},
\end{equation}
which are subject to the so-called \emph{betweenness condition},
\begin{align}
\label{eq:betweenness}
m_{k,l} &\geq m_{k,l-1} \geq m_{k+1,l} & (1 \leq k < l \leq N).
\end{align}
Tab.~\ref{tab:conversion} gives examples of equivalent Young tableaux and \GTP{}s.
\begin{table}
\begin{center}
\begin{tabular}{c@{$\qquad$}c@{$\qquad$}c@{$\qquad$}c}
\hline \hline \rule[-4ex]{0pt}{9ex}
{\tiny$\begin{pmatrix}
\multicolumn{2}{c}{2}
\end{pmatrix}$} &
{\tiny$\begin{pmatrix}
\multicolumn{2}{c}{3} & \multicolumn{2}{c}{2} \\
\hline
& \multicolumn{2}{c}{2} &
\end{pmatrix}$} &
{\tiny$\begin{pmatrix}
\multicolumn{2}{c}{3} & \multicolumn{2}{c}{2} & \multicolumn{2}{c}{1} \\
\hline
& \multicolumn{2}{c}{3} & \multicolumn{2}{c}{2} & \\
&& \multicolumn{2}{c}{2} &&
\end{pmatrix}$} &
{\tiny$\begin{pmatrix}
\multicolumn{2}{c}{4} & \multicolumn{2}{c}{3} & \multicolumn{2}{c}{1} & \multicolumn{2}{c}{1} \\
\hline
& \multicolumn{2}{c}{3} & \multicolumn{2}{c}{2} & \multicolumn{2}{c}{1} & \\
&& \multicolumn{2}{c}{3} & \multicolumn{2}{c}{2} && \\
&&& \multicolumn{2}{c}{2} &&&
\end{pmatrix}$}
\\ \hline \rule[-6ex]{0pt}{13ex}
$\young(11)$ & $\young(112,22)$ & $\young(112,22,3)$ & $\young(1124,224,3,4)$
\\ \hline \hline
\end{tabular}
\end{center}
\caption{Examples of \GTP{}s and corresponding Young tableaux;
the top row of each pattern determines the shape of its respective tableau.
These examples have been constructed in such a way that each tableau/pattern
contains the examples to its left as a sub-tableau/-pattern.
Note that we usually drop columns of \SU N Young tableaux with length $N$
(e.g., the leftmost column in the rightmost tableau);
this corresponds to subtracting the entry $m_{N,N}$ from each entry of a \GTP{}.
The weight of a state can be directly constructed from a \GTP{}:
Let $\sigma_l = \sum_{k=1,\dotsc,l} m_{k,l}$ denote the row sums,
then $\lambda^{M}_l = (\sigma_{l+1}-\sigma_l)/2$.}
\label{tab:conversion}
\end{table}

Young tableaux and \GTP{}s are composite labels, $M=(\Gamma,\gamma)$ in short,
in the sense that they can play  the role
of both the irrep label $\Gamma$ and the state label $\gamma$.
The shape of a Young tableau (i.e., without the labeling of boxes)
determines an irrep $\Gamma$; it corresponds to the top row of a \GTP{},
whose $k$-th entry $m_{k,N}$ specifies the number of boxes
in the $k$-th row of the Young tableau.
The dimension of an irrep $\Gamma$ is equal to the number of valid
\GTP{}s with a given top row. There exists a convenient formula for this number:
\begin{equation}
\label{eq:dimension}
\dim(\Gamma) = \prod_{1 \leq k < k' \leq N} \left( 1 + \frac{m_{k,N} - m_{k',N}}{k' - k} \right).
\end{equation}
 The GT labeling scheme has the advantage that 
for any of the generators $\mathcal{\hat J}^a \in 
\{\mathcal{\hat J}^{(l)}_{z,\pm}\}$, 
the corresponding matrix elements $\Gamma^a_{\gamma',\gamma}$ 
   [Eq.~(\ref{eq:multiplet_transformation})]
  within the irrep $\Gamma$ are known explicitly, given by a
  complicated formula worked out by Gelfand and
  Tsetlin.~\cite{GelfandTsetlin}

A further ingredient to \SU N representation theory is
the decomposition of a tensor product $\Gamma_1 \otimes \Gamma_2$ of two irreps
into a direct sum of irreps (see Eq.~\eqref{eq:product_decomposition}).
For \SU N, this trick is accomplished by the \emph{Littlewood-Richardson rule},
which is beyond the scope of this introduction.
It produces equations such as
\begin{equation}
\label{eq:littlewood_richardson}
\begin{split}
{\tiny\yng(3,1)} \otimes {\tiny\yng(2,1)} &= {\tiny\yng(1)} \oplus {\tiny\yng(2,2)} \oplus 2\:{\tiny\yng(3,1)} \\
&\quad \oplus {\tiny\yng(4)} \oplus {\tiny\yng(4,3)} \oplus {\tiny\yng(5,2)}.
\end{split}
\end{equation}
The number of times a particular irrep occurs on the right-hand side
is called its \emph{outer multiplicity}; for \SU N, it is $>1$ in
general.   The particular basis transformation effecting this
  decomposition is described by \emph{Clebsch-Gordan coefficients}
  [see Eq.~\eqref{eq:decomposition1}]; Ref.~\onlinecite{ArneJan}
presents a numerical algorithm for computing them for any $N$.

Now consider a quantum chain model involving
$N$ creation and annihilation operators per site,
$f_\alpha^{[n]\dagger}$ and $f_\alpha^{[n]}$ for site $n$, 
with $\alpha = 1, \dots, N$.
For a given site $n$, consider the set of operators 
\begin{equation}
\hat J^{a,[n]}  \equiv \sum_{\alpha,\beta=1}^{N}
{f^{[n]}_\alpha }^\dagger(J^a)_{\alpha,\beta} f^{[n]}_\beta \; , 
\label{eq:site_generators}
\end{equation}
with the matrices $J^a$ taken as the defining representation of \SU N
[Eq.~(\ref{eq:defining_representation})].
The $\hat J^{a,[n]}$ satisfy the same commutation
relations as the \SU N generators $\mathcal{\hat J}^a$ and hence
generate an operator representation of \SU N on the Fock space of site
$n$. The action of these operators partitions this Fock space into a
direct sum of irreps.  For example, Tab.~\ref{table:states} specifies
these irreps explicitly for the case of \SU 3.

The Fock space of a the full chain is the direct product of
  the Fock spaces of each site. Correspondingly, $\hat J^a =
  \sum_{\oplus n} \hat J^{a, [n]}$ generates an operator representation
  of \SU N on the, which can be decomposed 
  into a direct sum of irreps by iterative use of the
  Littlewood-Richardson rule and Clebsch-Gordan coefficients.

  The Hamiltonian $\hat H$ for the full chain has \SU N symmetry if it
  commutes with all generators $\hat J^a$. When the Hamiltonian is
  expressed in the Fock space basis just
  mentioned, consisting of a direct sum of \SU N irreps, it is block-diagonal,
   with each block containing matrix elements only between states transforming
  according to a given \SU N irrep. 

Diagonalizing such blocks, or more generally, calculating
matrix elements of operators, is expedited by using the 
Wigner-Eckart theorem. To this end, 
one needs to group
  operators in Fock space into operator multiplets (sometimes called
  \emph{irreducible tensor operators}). An operator
multiplet transforming according to the irrep $\Gamma$ 
is a set of operators $\hat
  O_{\Gamma, \gamma}$ that satisfy the relations
\begin{equation}
\left[\hat J^\alpha, \hat O_{\Gamma, \gamma}\right] = \sum_{\gamma'} \Gamma^\alpha_{\gamma', \gamma} \hat O_{\Gamma, \gamma'} .
\label{eq:tensorop}
\end{equation}
For example, the set of all generators ${\hat J}^\alpha$
  constructed in Eq.~\eqref{eq:site_generators} spans an operator
  multiplet, acting only on site $n$ and transforming according to the
  adjunct representation of \SU N, which has dimension $N^2-1$. (For
  \SU 3, this is the irrep $\tiny\yng(2,1)$ listed in
  Tab.~\ref{table:tensors}).  

A convenient way to explicitly construct an operator multiplet
associated with a given site $n$ is to find
  its highest-weight operator by guessing or repeated application of
  raising operators, and then produce the other operators in the
  multiplet by applying lowering operators, using
  Eq.~\eqref{eq:tensorop}.  Tab.~\ref{table:tensors}
gives some examples of \SU 3 operator multiplets constructed in this manner.

The tensor product of two operator multiplets, each acting on
individually on a separate site, can be decomposed into a direct sum of
two-site operator multiplets, again using the
  Richardson-Littlewood rule and Clebsch-Gordan coefficients.
However, one often finds that extending a single-site operator
multiplet to two sites by taking its tensor product with the identity
operator (transforming according to the trivial \SU N representation)
is enough, so a complicated decomposition can be avoided in most
cases.  For more than two sites, this procedure is applied
iteratively.

\begin{table}[tb]
\centering
\begin{tabular}{c|c|cc|c|c }
\hline
\hline
& & &      &                               &  \\
  Multiplets & States &       $Q$                       &    $ F$ & Energy & Degeneracy\\
& & &      &                               &  \\                                 
\hline   
& & &      &                               &  \\
1& $|1\rangle =  |0 \rangle $    & $-3/2$             &  $\bullet$ & 0 & 1 \\
& & &      &                               &  \\
& & &      &                               &  \\
2 & $ \left .
\begin{array}{c}
	|2 \rangle =  d_1^\dagger|0 \rangle \\
	\\
	|3 \rangle =  d_2^\dagger|0 \rangle \\
	\\
	|4 \rangle = d_3^\dagger|0 \rangle
\end{array} 
\right \} $
     &   $-1/2$           &  \yng(1) &  $\e_d$ & 3 \\
& & & & &  \\
& & & & &  \\
3 & $ \left .
\begin{array}{c}
	|5 \rangle =  d_1^\dagger d_2^\dagger|0 \rangle \\
	\\
	|6 \rangle =  d_1^\dagger d_3^\dagger|0 \rangle \\
	\\
	|7 \rangle = d_2^\dagger d_3^\dagger|0 \rangle 
	\end{array} 
\right \} $
     &   $1/2$           &  \yng(1,1) & $2\e_d +U$  &3 \\
& & & & & \\
& & & & & \\
4 &$|8 \rangle =  d_1^\dagger d_2^\dagger d_3^\dagger|0 \rangle $       &   $1/2$           &  $\bullet$ & $3\e_d+3 U $ & 1\\ 
& & & & & \\
\hline \hline
\end{tabular}
\caption{ Organization of the impurity states into multiplets. 
The $2^3=8$ states are organized into 4 multiplets, each 
characterized by a set of quantum numbers  $(Q, F)$. 
States within a multiplet are degenerate in energy.}
\label{table:states}
\end{table}

\section{Multiplets within the Anderson model with $U_Q(1)\times SU_F(3)$ symmetry}

In this appendix we illustrate in more detail how the general 
concepts presented in Appendix \ref{appendix:SU3} can be applied to our case,  when 
$U_Q(1)\times SU_F(3)$ symmetry is used. 
We first construct the lowering/raising operators in the Fock space 
by using   \eqref{eq:site_generators}.  Explicitly, for \SU 3  we shall need explicitly six generators,   
 \begin{eqnarray}
&&\hat J_+^{(1)}  =   d_{1}^{\dagger}d_{2},\quad\hat J_+^{(2)}  =   d_{2}^{\dagger}d_{3},
%\hat J_+^{(l)} & = &  d_{l}^{\dagger}d_{l+1}\quad (l=1,2)
\nonumber \\
&&\hat J_-^{(1)}  =   d_{2}^{\dagger}d_{1},\quad\hat J_-^{(2)}  =   d_{3}^{\dagger}d_{2},
%\hat J_+^{(l)} & = &  d_{l}^{\dagger}d_{l+1}\quad (l=1,2)
\nonumber \\
%\hat J_-^{(l)} & = &  d_{l+1}^{\dagger}d_{l}\\
\hat J_z^{(1)} & = & {1\over 2} \left ( d_{1}^{\dagger}d_{1}-d_{2}^{\dagger}d_{2}\right),
\hskip0.3cm
\hat J_z^{(2)}  =  {1\over 2} \left ( d_{2}^{\dagger}d_{2}-d_{3}^{\dagger}d_{3}\right).
\nonumber
%\\
%\hat J_z^{(l)} & = & {1\over 2} \left ( d_{l}^{\dagger}d_{l}-d_{l+1}^{\dagger}d_{l
%+1}\right)
\end{eqnarray}

The initial impurity states can be constructed and  organized into 4 \SU 3 multiplets relatively easily. 
They are presented in Table \ref{table:states}. The highest weight states $|\G, \g=\G\rangle$ 
can be found by requiring that both $\hat J_+^{(1)} $ and 
 $\hat J_+^{(2)} $ annihilate them. Further states within the multiplet are then obtained by acting with $\hat J_-^{(1)} $  and  $\hat J_-^{(2)}$, and comparing to  Eq.~\ref{eq:multiplet_transformation}.
Notice that the Clebsch-Gordan coefficients do depend on the particular choice of basis, and therefore, to make use of the Wigner-Eckart theorem, every multiplet must be constructed  to conform with the same choice of basis as the Clebsch-Gordan coefficients. This choice of basis 
is implicitly contained in the matrix elements ($\G^{(l)}_{\pm})_{\g\g'}$. In the open access Flexible NRG code~\cite{OpenAccess} we used the conventions of Ref.~\onlinecite{ArneJan}, for which these
matrix elements are explicitly given in  Ref.~\onlinecite{GelfandTsetlin}.
%{\gergely Remark: It would be useful to include these in the web page you constructed with Arne, and give a reference to it.} 

To use the Wigner-Eckart theorem, we need to 
organize operators  into operator multiplets.  
A few examples of these were presented in Tab.~\ref{table:tensors}. 
Let us discuss here the specific case 
of the spin operator,  forming a 8-dimensional multiplet, which transforms according to the 
$\tiny \yng(2,1)$ representation. A similar procedure applies  to the 
hopping or  the superconducting operators. 

We can quickly guess that the highest
weight operator is $d_1^\dagger d_3$ since it commutes  with both $\hat J_+^{(1)}$
and $\hat J_+^{(2)}$ (see Eq.~\eqref{eq:tensorop}).
 This operator thus corresponds to the  $\tiny \young(11,2)$ state. 
Now, by applying the lowering 
operators (i.e., forming commutators with them)  
as indicated in Fig. \ref{fig:weight-diagram}, we can generate all the 
other operators that form the multiplet.\cite{ArneJan} In this procedure one needs 
again the explicit form of the corresponding matrix elements, ($\G^{(l)}_{\pm})_{\g\g'}$ to fix the proper  
phase/sign of the basis states and their normalization. 

 We remark that particular care must be taken  in cases where a weight has an inner multiplicity, as  for the members 
$\tiny \young(13,2)$ and $\tiny \young(12,3)$ of the spin operator multiplet. There one needs to take the 
correct linear combination of the corresponding  operators to transform according to  ($\G^{(l)}_{\pm})_{\g\g'}$ (see Eq.~\eqref{eq:tensorop}).

\section{Recursion formulas}\label{recursion_formulas}

In this section we shall detail how the recursive relations 
(Eqs.~\eqref{eq:hopping}-\eqref{eq:local_operator})  were derived.  
The general procedure is based on the Wigner-Eckart 
theorem and using sum-rules satisfied by the Clebsch-Gordan coefficients.  
We first derive the recursion relation 
for the irreducible matrix element of the hopping Hamiltonian. 
The general expression of the hopping  matrix element at iteration 
$n$ reads
\begin{widetext}
\bea
_{\alpha_u} \left <u, \G_u, \g_u   \right | \tau_{n-1,n}   \left | v, \G_v, \g_v 
\right>_{\alpha_v}^{[n]}&  = &  \sum_{a}\left [ h_{a}^{[n-1]}\sum_{\g_a}\; 
 _{\alpha_u}\left <u, \G_u, \g_u   \right |  C_{a,\G_a,\g_a}^{[n-1]}(C_{a,\G_a, 
 \g_a}^{[n]})^\dagger  \left | v, \G_v, \g_v \right>_{\alpha_v}^{[n]}  +h.c. \right ]
 \label{eq:meh}\;. 
\eea
\end{widetext}
  Here the multiplets, $ \left |u, \G_u, 
\g_u   \right >_{\alpha_i}^{[n]}$ and $\left | v, \G_v, \g_v \right>_{\alpha_v}
^{[n]} $ were constructed  in 
terms of the block multiplet $ \left |i, \G_i, \g_i   \right >^{[n-1]}$
and local multiplet $ \left |\mu, \G_\mu^{\rm loc}, \g_\mu^{\rm loc}   \right >$
at iteration $n-1$ using Eq.~\eqref{eq:decomposition1}.
Next we exploit the locality of the $ C_{a,\G_a,
\g_a}^{[n]}$, i.e. that the  operator $ C_{a,\G_a,
\g_a}^{[n]}$ acts only on  local states $\mu$, while $ (C^{[n-1]}_{a,\G_a,\g_a})^{\dagger}$ acts 
 on the block states $i$.  The matrix element of Eq.~\eqref{eq:meh} then becomes
 \begin{widetext}
%%%%%%%%%%%%%%%%%%
\bea
_{\alpha_u} \left <u, \G_u, \g_u   \right | \tau_{n-1,n}   \left | v, \G_v, \g_v 
\right>_{\alpha_v}^{[n]}
 & = &  \sum_{a}\left [ h_{a}^{[n-1]}\; {\rm sgn}(C_{a}, \mu)\;   \sum_{\g_a}\; \sum_{\g_i,\g_\mu^{\rm loc}, \g_j \g_\nu^{\rm loc} }
\left ( \G_\mu^{\rm loc},\g_\mu^{\rm loc}; \G_i ,\g_i \mid  \G_u,\g_u   \right )_{\alpha_u}^{*}\times\right .\\
&  & \left ( \G_\nu^{\rm loc},\g_\nu^{\rm loc}; \G_j ,\g_j \mid  \G_u,\g_u   \right )_{\alpha_v}  \left < i, \G_i,\g_i   \right |  C_{a,\G_a,\g_a}^{[n-1]}  \left |  j, \G_j,\g_j     \right > ^{[n-1]}\nonumber \times\\
&  & \left .
\left <\nu, \G_\nu^{\rm loc}, \g_\nu^{\rm loc}  \right |  C_{a,\G_a,\g_a}^{[n]} \left | \mu, \G_\mu^{\rm loc}, \g_\mu^{\rm loc}  \right >
 +h.c. \right ]\;. 
 \nonumber
\eea
\end{widetext}
Here the sign function ${\rm sgn }(C_{a}, \mu)=\pm 1$ arises as we commute
the local state $\mu$ over the operator $C^{[n-1]}$. If the hopping  operator $C$ 
is fermionic and  the local state 
contains an odd number of fermions, then the sign is negative, otherwise it is positive. 
Now we can use the Wigner-Eckart theorem, Eq.~\eqref{eq:WE}, 
and express the matrix elements of the creation/annihilation operators in terms
of their reduced matrix elements. By doing that, we
immediately recover the result~\eqref{eq:hopping} with 
\begin{widetext}
\bea
D\left(a,\ua, \ub; u,v \right)& = &  
 {\rm sgn}(a, \mu)  \sum_{\ug_c}\sum_{\ug_i, \ug_\mu^{\rm loc}}\; \sum_{\ug_j,\ug_\nu^{\rm loc}}     
\left ( \uG_\mu^{\rm loc},\ug_\mu^{\rm loc}; \uG_i ,\ug_i \mid  \uG_{u},\ug_{u}   
\right )_{\ua_{u}}^{\ast}
\left ( \uG_\nu^{\rm loc},\ug_\nu^{\rm loc}; \uG_j ,\ug_j \mid  \uG_{u},\ug_{u}   
\right )_{\ua_{v}}\times 
\nonumber\\
& &  \left ( \uG_a,\ug_a; \uG_j ,\ug_j \mid     \uG_i,\ug_i   \right )_{\ua}^{\ast}
\left ( \uG_a,\ug_a; \uG_\mu^{\rm loc} ,
\ug_\mu^{\rm loc} \mid  \uG_\nu^{\rm loc},\ug_\nu^{\rm loc}  \right )_{\ub} .
\label{eq:D}
\eea

A slightly different analysis can be done for block operators. 
First we notice that we can ``invert'' the Wigner-Eckart theorem, Eq.~\eqref{eq:WE},
using the completeness of the Clebsch-Gordan coefficients,
and express the reduced matrix elements instead as:
\bea
\left < u\parallel A\parallel v \right >_{\beta} ^{[n]}
& =& 
\sum_{\g_A, \g_v}\, _{\alpha_u}\left <u, \G_u, \g_u   \right | A_{\G_A, \g_A}   
\left | v, \G_v, \g_v \right>_{\alpha_v} ^{[n]}
\left ( \G_A,\g_A; \G_v ,\g_v \mid  \G_u,\g_u   \right )_{\beta}.
\label{eq:irreducible_local}
\eea
\end{widetext}
As a next step,  we need to express the matrix element $ _{\alpha_u} \left <u, \G_u, \g_u   \right | A_{\G_A, \g_A}    
\left | v, \G_v, \g_v \right>_{\alpha_v}^{[n]} $ in
\eqref{eq:irreducible_local} in terms of the irreducible matrix elements
of the operator at iteration $n-1$. Here we follow the same strategy as in the 
case of the hopping operator: We expand first the states using
\eqref{eq:decomposition}, while keeping in mind that the operator acts only in 
the block states sector, and then use the 
Wigner-Eckart theorem~\eqref{eq:WE} for the 
operator's matrix elements. By doing that we find:
\begin{widetext}
\bea
 _{\alpha_u} \left <u, \G_u, \g_u   \right | A_{\G_A, \g_A}    
\left | v, \G_v, \g_v \right>_{\alpha_v}^{[n]} 
& = &  \sum_{\g_i, \g_j}\sum_{\g_\nu^{\rm loc},\g_\mu^{\rm loc}}\; \delta_{\mu,
\nu}\delta_{\G_\mu^{\rm loc}, \G_\nu^{\rm loc}}\delta_{\g_\mu^{\rm loc},\g_
\nu^{\rm loc}}
\left ( \G_\mu^{\rm loc},\g_\mu^{\rm loc}; \G_i ,\g_i \mid  \G_u,\g_u   
\right )_{\alpha_u}^*\nonumber
\\
&  & 
\left ( \G_\nu^{\rm loc},\g_\nu^{\rm loc}; \G_j ,\g_j \mid  \G_v,\g_v   
\right )_{\alpha_v}\times \nonumber \\
& & {\rm sgn}(A,\mu) \sum_{\alpha}
\left ( \G_A,\g_A; \G_j ,\g_j \mid  \G_i,\g_i   \right )_{\alpha}^* \left < 
i \parallel A\parallel j \right >_{\alpha}^{[n-1]}.
\label{eq:matrix_element}
\eea
Plugging Eq.~\eqref{eq:matrix_element} in Eq.~\eqref{eq:irreducible_local}
we recover the result for the block operator stated in \eqref{eq:block_operator} with the coefficient $F$ defined as
\bea
 F\left (\ua, \ub; u, v \right ) &  = &  
 {\rm sgn}(A,\mu)\;\sum_{\ug_\mu^{\rm loc}}\sum_{\ug_i,\ug_j}\sum_
{\ug_A,\ug_{\tj}}
\left ( \uG_\mu^{\rm loc},\ug_\mu^{\rm loc}; \uG_i ,\ug_i \mid  \uG_{u},\ug_{u}   \right )_{\ua_{u}}^{\ast} 
\left ( \uG_\mu^{\rm loc},\ug_\mu^{\rm loc}; \uG_j,\ug_j \mid  \uG_{v},\ug_{v}   \right )_{\ua_{v}} \times 
\nonumber\\
& & \left ( \uG_A,\ug_A; \uG_j,\ug_j \mid  \uG_i,\ug_i   \right )_{\ua}^{\ast} 
\left ( \uG_A,\ug_A; \uG_{v},\ug_{v} \mid  \uG_{u},\ug_{u}   \right )_{\ub}.
\eea

A similar analysis can be done in the case of a ``local operator'',
giving the final expression for the coefficient $K$
entering Eq.~\eqref{eq:local_operator}
\bea
K\left (\ua, \ub; u,v \right ) &  = &   
\sum_{\ug_i}\sum_{\ug_\mu^{\rm loc},\ug_\nu^{\rm loc}}\sum_{\ug_A,\ug_{v}}
\left ( \uG_\mu^{\rm loc},\ug_\mu^{\rm loc}; \uG_i ,\ug_i \mid  \uG_{u},\ug_{u}   
\right )_{\ua_{u}}
^{\ast} \left ( \uG_
\nu^{\rm loc},\ug_\nu^{\rm loc}; \uG_i,\ug_i \mid  \uG_{v},\ug_{v}   
\right )_{\ua_{v}} \times \nonumber\\
& & \left ( \uG_A,\ug_A; \uG_\nu^{\rm loc},\ug_\nu^{\rm loc} \mid  \uG_\mu^{\rm 
loc},\ug_\mu^{\rm loc}   \right )_{\ua}^{\ast} 
\left 
( \uG_A,\ug_A; \uG_{v},\ug_{v} \mid  \uG_{u},\ug_{u}   \right )_{\ub}  .
\eea
\end{widetext}

\section{Connection to the Matrix Product States approach}
\label{append:MPS}

It is well-known that the states constructed within the NRG
framework can be viewed as Matrix Product states (MPS)\cite{Weichselbaum2009}.
Moreover, it has been shown recently 
that non-Abelian symmetries can be incorporated into 
the construction of MPS\cite{Singh2010,Andreas}. 
In this appendix we briefly review how this can be
done in the context of NRG.
We  start with the simple observation that the Hamiltonian of a chain of length $N$, $H_N$
%\begin{equation}
%{H_N} = {\cal H}_{0} +\sum_{n=0}^{N-1}\bigl( \tau_{n, n+1} +{\cal H}_{n+1}\bigr) \;, 
%\label{eq:H_N}
%\end{equation}
%!TEX encoding = UTF-8 Unicode
%
acts on the Hilbert space spanned by a basis constructed from local 
multiplets: 
\begin{equation}
\left \{
| \mu_0, \G_{\mu_0}^{\rm loc}, \g_{\mu_0}^{\rm loc}; 
\mu_1, \G_{\mu_1}^{\rm loc}, \g_{\mu_1}^{\rm loc};
\dots;
\mu_N, \G_{\mu_N}^{\rm loc}, \g_{\mu_N}^{\rm loc} 
\rangle 
\right \}\;.
\end{equation}
The dimension of this basis set is $d_0 d^N$, with $d_0$  the dimension
 of ${\cal H}_{0}$, and  $d$ is the dimension of the Hilbert space at sites
along the Wilson chain (see Tab.~\ref{table:states}). For the \SU 3 Anderson model we have four multiplets on each site, and $d_0=d=8$.  

Let us now assume that we have constructed 
somehow some block  states $ i$, which span the relevant part of the Hilbert space 
of a chain with $n-1$ sites. The number of these states, $D$ is, of course, much less 
than the total number of states within this block, which would be of the order of 
$\sim d^{n}$. We can then use these states to express the appropriate (relevant) states 
for a chain of  $n$  sites as 
\begin{multline}
| \ti, \G_\ti, \g_\ti\rangle^{[n]}_{\alpha_\ti} = 
\sum_{i, \G_i, \g_i}^{\sim}
\sum_{\mu, \G_{\mu}^{[\rm loc]}, \g_{\mu}^{[loc]}}
\left (  
P^{[\G_{\mu}^{\rm loc} \g_{\mu}^{\rm loc}]}_{\G_i\g_i, \G_\ti \g_\ti; \alpha_\ti } \right)^{[\mu]}_{i\ti}\;\\
|\mu, \G_\mu^{\rm loc},\g_\mu^{\rm loc}\rangle\otimes |i, \G_i,\g_i \rangle^{[n-1]}\;.
\label{eq:decomposition}
\end{multline}
Here $P$ is some projector that generates the relevant block of 
multiplets $\{|\ti\rangle\}$ from the block multiplets $\{|i \rangle\}$ by adding 
some local states $\{ |\mu\rangle\}$. The tilde in the sum indicates that only
a number $D$ of multiplets are 
kept at each iteration. In the presence of symmetries, the matrices $P$ can be 
factorized into products of reduced matrix elements
and Clebsch-Gordan coefficients, as:
\begin{equation}
\left ( P^{[\G_{\mu}^{\rm loc} \g_{\mu}^{\rm loc}]}_{\G_i\g_i, \G_\ti \g_\ti; \alpha_\ti } \right)^{[\mu]}_{i\ti}\;
= \left (A^{[\G_{\mu}^{\rm loc}]}_{\G_i\G_\ti } \right)^{[\mu]}_{i\ti}\,
\left ( C^{[\G_{\mu}^{\rm loc}]}_{\G_i \G_\ti; \alpha_\ti } \right)^{ [\g_{\mu}^{\rm loc}]}_{\g_i\g_\ti}\;, 
\label{eq:factorization}
\end{equation}

with
\begin{equation}
\left ( C^{[\G_{\mu}^{\rm loc}]}_{\G_i \G_\ti; \alpha_\ti } \right)^{[\g_{\mu}^{\rm loc}]}_{\g_i\g_\ti}\; = 
\left( \G_i,\g_i; \G_{\mu}^{\rm loc},\g_{\mu}^{\rm loc} \mid  \G_{\ti},\g_{\ti}   \right )_{\alpha_\ti}.
\label{eq:factorization}
\end{equation}

Here the key observation is that  the reduced matrix elements $ (A^{[\G_{\mu}^{\rm loc}]}_{\G_i\G_\ti } )^{[\mu]}_{i\ti}$
can also be thought of as the matrix elements of some irreducible operators,  
 labeled by $\G_{\mu}^{\rm loc}$.

Repeating the iteration procedure multiplets describing the full chain of size 
$N$ can be constructed this way as a matrix product state 
(summation of repeated indices is implicitly assumed)

\begin{widetext}
\begin{multline}
%\begin{eqnarray}
|\psi\rangle^{[N]}=| i_\psi, \G_{i_\psi}, \g_{i_\psi}\rangle^{[N]} 
%& = & 
=
\left (A^{[\G_{\mu_0}^{\rm loc}, \g_{\mu_0}^{\rm loc}]}_{\G_{i_0} } \right)^{[\mu_0]}_{i_0}\;
\left (A^{[\G_{\mu_1}^{\rm loc}]}_{\G_{i_0}\G_{i_1} } \right)^{[\mu_1]}_{i_0 i_1}\;
\left ( C^{[\G_{\mu_1}^{\rm loc}]}_{\G_{i_0} \G_{i_1}; \alpha_{i_1} } \right)^{[\g_{\mu_1}^{\rm loc}]}_{\g_{i_0}\g_{i_1}}\;
%\times\nonumber\\&  & 
\left (A^{[\G_{\mu_2}^{\rm loc}]}_{\G_{i_1}\G_{i_2} } \right)^{[\mu_2]}_{i_1 i_2}\;
\left ( C^{[\G_{\mu_2}^{\rm loc}]}_{\G_{i_1} \G_{i_2}; \alpha_{i_2} } \right)^{[\g_{\mu_2}^{\rm loc}]}_{\g_{i_1}\g_{i_2}}\;
%\times  
\nonumber\\
%&  & 
\times \dots\times\left (A^{[\G_{\mu_{N}}^{\rm loc}]}_{\G_{i_{N-1}}\G_{i_\psi} } \right)^{[\mu_N]}_{i_{N-1} i_\psi}\;
\left ( C^{[\G_{\mu_{N}}^{\rm loc}]}_{\G_{i_{N-1}} \G_{i_{\psi}}; \alpha_{i_{\psi}} } \right)^{[\g_{\mu_{N}}^{\rm loc}]}_{\g_{i_{N-1}}\g_{i_{\psi}}}\;
%\times \nonumber\\
%& & 
| \mu_0, \G_{\mu_0}^{\rm loc}, \g_{\mu_0}^{\rm loc}; 
\mu_1, \G_{\mu_1}^{\rm loc}, \g_{\mu_1}^{\rm loc};
\dots;
\mu_N, \G_{\mu_N}^{\rm loc}, \g_{\mu_N}^{\rm loc} 
\rangle \;.
\label{eq:MPS_state}
\end{multline}
%\end{eqnarray}
\end{widetext}

%In Eq.~\eqref{eq:MPS_state}, we can view
Note that the index structure that arises here implies matrix multiplication
not only for the $A$-matrices of reduced matrix elements, but also for the $C$-matrices of Clebsch-Gordan coefficients.
This MPS formulation can be used either to implement standard Wilsonian truncation
(as done in Ref.~\onlinecite{Andreas}), or, alternatively,
to proceed variationally, as done in the density-matrix renormalization
group (DMRG),\cite{Schollwock2011} whose use for 
Wilson chains was explored in \cite{Saberi2008,Weichselbaum2009}.
In the latter case, one views
the $A$ matrices as a set of variational 
parameters that need to be optimized according to some criteria.
To find the optimal approximation for the ground state, e.g.,  we look for 
the corresponding MPS which minimizes the total energy,
\begin{equation}
E = \frac{\langle \psi |H_N| \psi\rangle ^{[N]} }{ \langle \psi | \psi\rangle ^{[N]}}\;.
\end{equation}
This variational problem can be converted into a generalized eigenvalue problem 
and  solved using an iterative sweeping-like procedure 
% and by introducing  some Lagrange multiplier which turn it into
(see   [\onlinecite{Schollwock2011}] for more technical details). Once the MPS state 
is found,  it is possible to construct the eigenspectrum of the Hamiltonian 
$H_N$ for a fixed $N$. The flow diagram, such as the one presented in 
Fig.~\ref{fig:flow_diagram_half_filling}, can be obtained by tracing the 
spectrum of the Hamiltonian with increasing $N$; if
the discretization parameter $\Lambda$ is large enough ($\Lambda \gtrsim 2$),
the numerical results thus obtained are essentially equivalent
to those using standard Wilsonian truncation\cite{Saberi2008}.  For more details on
how the operators can be treated at the MPS level we direct the reader to the 
more detailed reviews, Ref.~\onlinecite{Schollwock2011,Andreas}.


\begin{thebibliography}{99}
\bibitem{typos} Notice also that some equations in
 Ref.~\onlinecite{Toth} contained typos, which we tried to eliminate
 here.
\bibitem{footnote2}A local operator always carries an extra label indicating the position along the Wilson chain where it acts.
\bibitem{footnote3} Some care is needed with the usage of the Hermitian conjugation sign. In certain cases, one may call an operator multiplet, $A=F^\dagger$, e.g. In this case, 
$((F^\dagger)_{\g_F})^\dagger$ is not necessarily equal to $F_{\g_F}$.
\bibitem{footnote_charge} As Q is semi-integer in the 
code we have use $2Q$ as the charge quantum numbers.
\bibitem{footnote4}   {Notice that the Clebsch-Gordan coefficients are basis-dependent, and to use the Wigner-Eckart theorem, e.g., one always needs to make sure that the basis operators within an operator multiplet transform identically with the states which were used to compute the Clebsch-Gordan coefficients.}
\end{thebibliography}
\end{document}